\documentclass[preprint,12pt]{article}

\usepackage{authblk}
\usepackage{amssymb}
\usepackage{subcaption}
\usepackage{algorithm,algpseudocode}
\usepackage{dirtytalk} 
\usepackage[frozencache,cachedir=.,newfloat]{minted}
\setminted{fontsize=\small}
\usepackage{svg} 
\usepackage{hyperref}
\newenvironment{code}{\captionsetup{type=listing}}{}
\SetupFloatingEnvironment{listing}{name=Source Code}

\begin{document}

\title{Mìmir: A real-time interactive visualization library for CUDA programs}

\author[1]{Francisco Carter}
\author[1]{Nancy Hitschfeld}
\author[2]{Cristóbal A. Navarro}

\affil[1]{Department of Computer Science, University of Chile}
\affil[2]{
Institute of Informatics, Universidad Austral de Chile}

\maketitle

\begin{abstract}
Real-time visualization of computational simulations running over graphics processing units (GPU) is a valuable feature in modern science and technological research, as it allows researchers to visually assess the quality and correctness of their computational models during the simulation. Due to the high throughput involved in GPU-based simulations, classical visualization approaches such as ones based on copying to RAM or storage are not feasible anymore, as they imply large memory transfers between GPU and CPU at each moment, reducing both computational performance and interactivity. Implementing real-time visualizers for GPU simulation codes is a challenging task as it involves dealing with i) low-level integration of graphics APIs (e.g, OpenGL and Vulkan) into the general-purpose GPU code, ii) a careful and efficient handling of memory spaces and iii) finding a balance between rendering and computing as both need the GPU resources. In this work we present \texttt{Mìmir}, a CUDA/Vulkan interoperability C++ library that allows users to add real-time 2D/3D visualization to CUDA codes with low programming effort. With \texttt{Mìmir}, researchers can leverage state-of-the-art CUDA/Vulkan interoperability features without needing to invest time in learning the complex low-level technical aspects involved. Internally, \texttt{Mìmir} streamlines the interoperability mapping between CUDA device memory containing simulation data and Vulkan graphics resources, so that changes on the data are instantly reflected in the visualization. This abstraction scheme allows generating visualizations with minimal alteration over the original source code, needing only to replace the GPU memory allocation lines of the data to be visualized by the API calls provided by \texttt{Mìmir} among other optional changes. The studied benchmark visualizes an interop-mapped point cloud with displacement computed with a CUDA kernel, comparing performance against point cloud visualization that reads data from RAM. Results show \texttt{Mìmir} framerate is up to $9\times$ faster, and the total visualization time is up to $12\times$ faster, while using 1.5 times less GPU memory. 
\end{abstract}


\section{Introduction} \label{sec:intro}

Scientific computer visualization is a useful tool in various research fields, including computational physics, to communicate information in a concise and intuitive way that complements the standard experiment outcomes such as plots and tables. Data visualization can be used to enrich the presented results, but it also has utility during the research and development stage to detect errors or anomalies that would require extensive debugging otherwise (e.g. through assertions or prints in code). When real-time visualization is possible, visual inspection is valuable in computer simulations for analyzing physical phenomena occurring over time through a number of discrete time steps. Moreover, working with large and/or complex datasets requires some degree of interactivity to focus on features of interest contained in the data. Basic interactive capabilities come in the form of camera manipulation and changing visual attributes like color or lightning settings, while more advanced features may depend on the nature of the visualized experiment.

On the downside, generating visualizations that fulfill the above requirements is hard. The required processing pipeline may be highly heterogeneous due to different softwares, programming languages, libraries and file formats involved, which makes it difficult to package and document. This is the case for visualization tools implemented as applications and that lack bindings for programming languages, so the only possible entrypoint is loading data from file, which in turn requires the effort of organizing the experiment output in the specified format.

The complexity of the simulation code may further increase the inherent difficulty in producing real-time and/or interactive visualizations from it. A clear example of such complexity can be found in GPU-accelerated simulations, which are implemented using a non-trivial massive parallelism programming model, which by itself is considered challenging. These massively-parallel algorithms are typically implemented using NVIDIA's CUDA platform (OpenCL is another open source option), which is also the case in computational physics. Although GPU-parallel algorithms exhibit complex source codes, they are an interesting target for visualizations, as the simulation data already resides in the GPU's VRAM, ready to be potentially rendered to the screen without needing to travel to/from CPU's RAM. With graphics APIs such as OpenGL and Vulkan, this is possible, however they introduce their own complexities and low-level abstractions, specially Vulkan which is considered state-of-the-art in graphics. For these reasons, there is a need for a high-level library that allows real-time visualization of GPU-based simulations, without requiring the programmer to have knowledge in low-level 3D graphics APIs such as Vulkan.  

There are a variety of computational physics applications with common properties between them that could benefit from real-time visualizations. Physical simulation models can be grouped by their spatial domain type, dimensionality and the underlying computational data structures that hold up the simulations. Spatial domains can be structured; which is where element positions are implicitly coded on the represented structure, such as cells in finite difference methods (FDMs) or spins from a square lattice in the Potts model. For unstructured domains, element positions are free in space, and auxiliary structures such as space partition trees or graphs are often used to aid in the interactions within the elements, such as nearest neighbors search. Table \ref{tab:models-char} characterizes some well known physical models according to their spatial domain and computational data structures.

\begin{table}[ht!]
\resizebox{\columnwidth}{!}{\begin{tabular}{|l|l|l|}
\hline
\textbf{Physical model} & \textbf{Spatial domain type} & \textbf{Computational structures} \\ \hline
N-Body simulation & Unstructured 2D \cite{CARTER2018148} \& 3D \cite{bedorf2012sparse} & delaunay triangulation \cite{CARTER2018148}, octree \cite{bedorf2012sparse}\\ \hline
Molecular dynamics & Unstructured 3D & Verlet lists \cite{soma} \\ \hline
Cellular Potts model  & Unstructured 3D & Adaptive cell array \cite{TAPIA2011857} \\ \hline
Potts/Ising model & Structured 2D \cite{KOMURA20121155} \& 3D \cite{NAVARRO201648} & Cartesian grid \\ \hline
Cellular automata \cite{FERRANDO2011628} & Structured & Cartesian grid \\ \hline
Fluid dynamics (CFD) & Structured & Cartesian grid \cite{goodnight2007cuda} \\ \hline
Euler Tours & Structured 2D & de-Bruijn graph \cite{6062988} \\ \hline
\end{tabular}}
\caption{Characterization of physical models in terms of spatial domain.}
\label{tab:models-char}
\end{table}

There are few solutions that leverage the advantages identified in GPU algorithms to generate efficient visualizations. Traditional alternatives are oriented towards reading input datasets from RAM or disk, so to adjust for the GPU case, it is necessary to program device-to-host memory transfers and data serialization methods. For multiple state visualizations, the above process would be repeated once every iteration, degrading both the experiment and visualization performances. This scheme also scales poorly for visualization complexity, because adding a new visual property derived from data (e.g. particle color or size) means redoing the execution and updating the transfer/serialization routines that include the new property, for example, as a data column in the serialized file. 

The proposed Mìmir library implements a visualization pipeline using a Vulkan backend, which then can be called from experiment code written in CUDA. This allows computational physics scientists to leverage the benefits of state-of-the-art Vulkan graphics without incurring into its complex low-level API. It is worth mentioning that it is indeed possible to implement this workflow using a different combination of compute (the code running the experiment) and graphics (the code producing display) APIs, although for this work we have restricted to CUDA + Vulkan. 

Internally, the workflow for using a graphics API to visualize data in GPU memory processed by a compute API consists of establishing interoperatibility between the two APIs at the memory layer. Both compute and graphics APIs have read/write access to interoperable mapped memory, better known as \textit{interop} mapped memory. In terms of concurrency, a positive aspect in physical simulations is that only the compute API writes into the simulation data (i.e., the next state), while the graphics API only reads the latest state from that region of memory. This scenario, free of read/write race conditions, saves the code from adding critical regions or atomic operations that would otherwise produce more overhead. 

Interoperability with CUDA is possible from Vulkan, OpenGL and Direct3D, and it is also possible to interop between Vulkan and OpenGL as well. The methodology for establishing interop is well documented for all cases \cite{guide2013cuda}, but there are important implementation details for each one. CUDA / OpenGL interop requires mapping the OpenGL resource (e.g., buffer or texture) to a CUDA C/C++ (e.g., raw pointer or texture handle) before accessing to its contents inside a CUDA kernel. However, this mapping must be undone before using this resource from the graphics pipeline, in order to avoid undefined behavior. For physics simulations, this involves continuously re-mapping the interop resources while making sure it is not being used in rendering or computing operations. Multi-threading adds another layer of complexity, as the programmer also has to make sure the thread making interop calls also has the current OpenGL context.

For CUDA / Vulkan interoperability, mapped memory must be allocated through Vulkan as external memory, which then can be imported in CUDA and be treated as a regular memory pointer. An interop allocation mapped in CUDA / Vulkan is mapped to the same GPU memory region and can be used as a native allocation from the respective API without noticeable performance overhead. On the other hand, native CUDA allocations can coexist with interop memory without modifying kernel code reading or writing to it, and in the same way, the external interop memory property is transparent to Vulkan resources referencing it. Only memory needing to be accessed from both sides (for participating in visualization) needs to undergo the interop treatment.

These CUDA / Vulkan properties open the opportunity to use interop for real-time large scale visualization, as there is no memory transfer needed to setup the data into a location visible to the graphics backend. Having interop memory is a step forward real-time visualization of GPU data, but it does not end there. After the interoperability has been established, the graphics programming phase comes. Interop memory needs to be referenced in Vulkan graphics resources from a shader program, which in turn needs pipeline objects, descriptors and rendering related structures such as queues, command buffers and synchronization resources. All these components of the Vulkan programming model require profound graphics API programming knowledge that is out of the scope of parallel GPU-based simulation using CUDA. Having a library that could abstract these technical graphics aspects of real-time visualization, while at the same time leveraging state-of-the-art graphics pipelines such as Vulkan, would enable computational scientists study physical simulations more efficiently.

The rest of the manuscript is organized as follows: Section \ref{sec:related} covers related works regarding visualization tools, Section \ref{sec:description} the technical details of the proposed Mìmir library, Section \ref{sec:results} presents experimental results of the library in terms of performance and overhead, and Section \ref{sec:discussion-conclusions} discusses the results and provides the main conclusions of this work.

\section{Related work} \label{sec:related}
There is a wide range of data visualization tools available today, but only a subset of them can be used to visualize and control an ongoing computational physical simulation properly. Three aspects can be evaluated when considering a library for these purposes; i) level of interactivity and real-time performance, ii) range of geometries supported, iii) ease of integration with existing simulation code. Most scientific visualization tools feature some degree of interactivity in the form of basic camera control, including those considered here as related work.

ParaView \cite{paraview} is a mature tool for interactive scientific visualization, with support for GPU acceleration for its VTK (The Visualization Toolkit) backend, and a client-server architecture for running remotely. Input for visualization comes from loaded files, and it is possible to generate animations either by loading a file series related by name. Likewise, it is also possible to generate animated output by recording a series of keyframes, each one with its own values for the various properties in the visualization (sources, filters, camera position, etc.). Other interactive capabilities include interactive or programmatic data selection. ParaView uses the VTK data model, which defines structured and unstructured grids (in terms of their topology being implicitly defined or not), meshes, tables and  pieced datasets. Discrete values can be defined over a mesh to form interpolated fields. Scripting for data analysis, selection and processing is made in Python, with support for viewing and loading scripts interactively through the API.

Datoviz \cite{datoviz} is an intermediate-level library, written in \texttt{C/C++} with Python bindings, that shows the feasibility of leveraging general purpose GPU programming and the state-of-the-art Vulkan API for scientific visualization. Its interface is based on abstractions for the application handle, display window, panels (or plots) and visualizations of different types, which can be further customized in terms of position, size and color, among others. Interactive support includes camera control and definition of custom callbacks for mouse and keyboard events. Geometry in Datoviz is organized in visuals and grouped by geometry, where 0D visual are pixels, points and glyphs in space, 1D visual represent lines, 2D visuals are images and 3D visuals are meshes, volumes and slices. Being a low-level library, it requires allocating and populating GPU memory directly through the library API, which can be accessed directly through the \texttt{C} interface or using the higher-level Python bindings.

The Open Visualization Tool Ovito \cite{ovito} is a multi-platform visualization software focused on particle simulation. It features an editable data processing pipeline where a dataset can pass through successive transformations before outputting the visualization, whose processed data can be exported in different formats. A visualization is composed of a number of visual elements of various types, including particles, vectors, or voxel grids in two and three dimensions. The tool has a programming interface in Python for data manipulation and rendering.

MegaMol \cite{megamol} is a framework originally focused on rendering particle data and visualization / analysis of molecular dynamics, with support for plugins such as mesh rendering, video recording and support for the OSPRay ray-tracing library. Its visualization is modeled as data sources (the dataset) and data sinks (the visual output). Within the library, Modules are programmable entities that define functionality in the form of data processing algorithms or a rendering pipeline. Modules can be interacted with through parameters, while Calls are used to link modules together in the form of data requests. This framework uses a mixed approach for linking render logic with input data. While modules for molecular dynamics use in-built shaders, mesh rendering requires a shader source code matching the vertex specification (attribute layout, shader input, etc.) applied to the mesh data.

VoroTop \cite{Lazar_2018} is focused on Voronoi cell and diagram visualization and analysis. Its interactive features include filtering cell topology families defined on plain text files loaded alongside the dataset when starting the tool. It supports both 2D and 3D geometries, although the tool is focused on 3D visualization with support for planar and profile views. This software features no programming language integration, but it is also available as an OVITO plugin.

\section{Introducing the Mìmir Library}
\label{sec:description}

The proposed solution consists in a library, named Mìmir, that implements a Vulkan graphics engine to generate visualizations from CUDA through interoperability between the two platforms. This engine is designed to allow CUDA programmers to visualize their simulation data with a state of the art API such as Vulkan, without incurring into the low-level complexities, while at the same time achieving efficient real-time performance. Internally, Mìmir initializes a Vulkan context for a CUDA-enabled GPU, allocates interop-mapped memory on the selected device, creates Vulkan graphics resources referencing the created memory, and renders a visualization from user-provided parameters. All of these tasks require a considerable amount of Vulkan code, whose function calls require numerous parameters that can produce crashes if chosen wrongly. Mimìr aims to streamline this setup process and only expose parameters relevant to the user in a scientific visualization context, avoiding low-level interaction with the Vulkan API.

\begin{figure}[H] \centering
	\includegraphics[width=\textwidth]{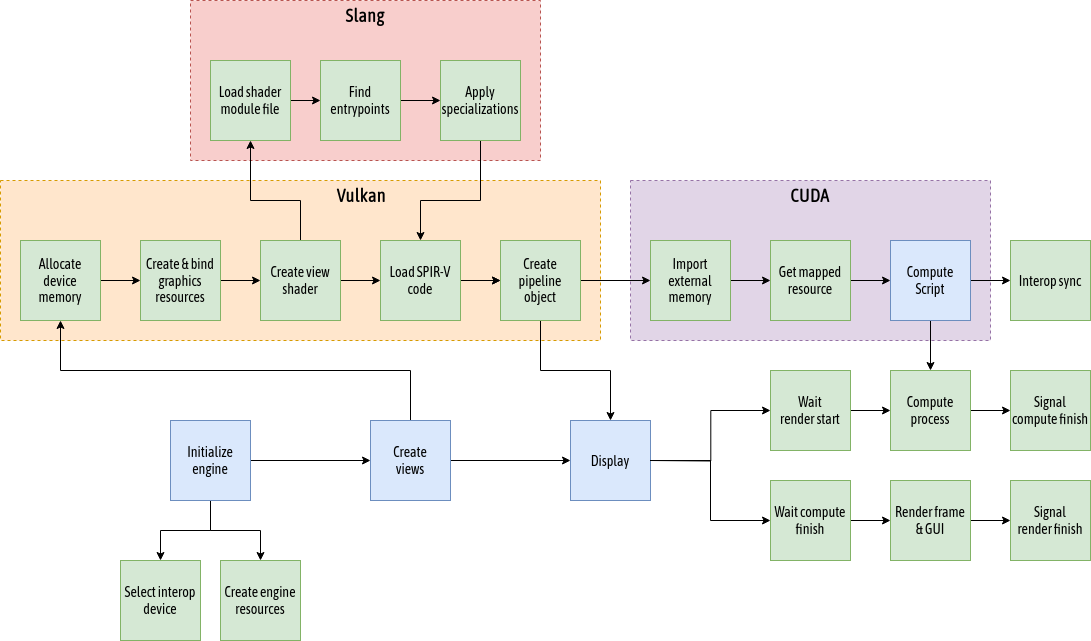}
	\caption{Overview of Mìmir library processes. Only blue stages are handled by the user, while green stages are managed by the library itself.}
	\label{fig:overview}
\end{figure}

Figure \ref{fig:overview} shows the proposed processing pipeline and its components, where processes in green are performed by the library with input received from the stages in blue. The library is designed to be used from CUDA code containing an experiment to be visualized, where allocations, kernel calls and cleanup are managed by the user as usual for a GPU-accelerated code. 

\subsection{Initialization}

The library visualization workflow is divided into three stages: i) instance initialization and creation of interop resources for their usage in experiment code, ii) creation of graphics resources required for rendering, and iii) interop synchronization management when rendering is running. Initialization requires to find and use an interop-capable device, that is, a GPU capable of running both CUDA and Vulkan. While it is possible to use multiple GPUs for computing (whose logic is entirely managed by the experiment code), the data to be visualized must reside in the interop device. Other initialization parameters include display options such as window configuration and target frame rate.

\subsection{Interface}

The library interface is designed to streamline the interoperability setup while encapsulating the Vulkan code from existing CUDA experiment code. Interaction between experiment code and the library is required at three stages: initializing the graphics engine and interop resources allocated from it, describing visualizations to define the requested visual output, and rendering. The initialization stage includes initializing the Vulkan instance, device and physical device handles, command pool and synchronization structures. Though some structures do not have a direct OpenGL counterpart, collectively they can be considered as equivalent to the OpenGL context. The visualization stage consists in providing all required information to correctly interpret interop data as one or possibly more visual components. The library treats inputs as raw memory regions and does not perform assumptions over them, which means users are required to provide explicit definition of all the parameters in the description. Nevertheless, these represent high-level visual interpretation options and technical knowledge about Vulkan is not required to correctly interpret them. Finally, the display stage optionally includes interop synchronization management between experiment code and the graphics backend in case the default scheme provided does not fit to a particular compute workflow, that is, the CUDA kernels or API functions writing over interop memory.

\subsubsection{Allocations}

The library manages device (GPU) memory allocations for CUDA / Vulkan interop, which require more setup than regular allocations made from either API. By design, Mìmir allocation functions use the same parameters as the native CUDA allocations shown in table \ref{tab:alloc-types}, with additional library-related inputs. These inputs include the handle to a previously created library instance, and the allocation handle that will contain the new allocation data. When successful, Mìmir allocations return the pointer to the CUDA resource as if it were created by the corresponding native function. Then, the corresponding allocation handle can be passed to library functions further down the visualization pipeline. 

\begin{table}[ht!]
\resizebox{\columnwidth}{!}{\begin{tabular}{|l|l|l|l|}
\hline
\textit{Memory type} & \textit{Mìmir function} & \textit{CUDA function} & \textit{Handle type} \\ \hline
Linear & \texttt{allocLinear} & \texttt{cudaMalloc} & \texttt{void*} \\ \hline
Opaque & \texttt{allocMipmap} & \texttt{cudaMallocMipmappedArray} & \texttt{cudaMipmappedArray} \\ \hline
\end{tabular}}
\caption{Mapping between CUDA allocation functions and Mìmir interop memory functions.}
\label{tab:alloc-types}
\end{table}

When using Mìmir in existing CUDA code, allocations that contain data involved directly in visualization need to be replaced by the appropriate library call as shown in table \ref{tab:alloc-types}. Other allocations do not need to be replaced even when used in kernels containing interop-mapped memory, as it behaves the same as regular memory under this setup.

\subsubsection{Views}

\begin{figure}[H] \centering
	\includegraphics[width=\textwidth]{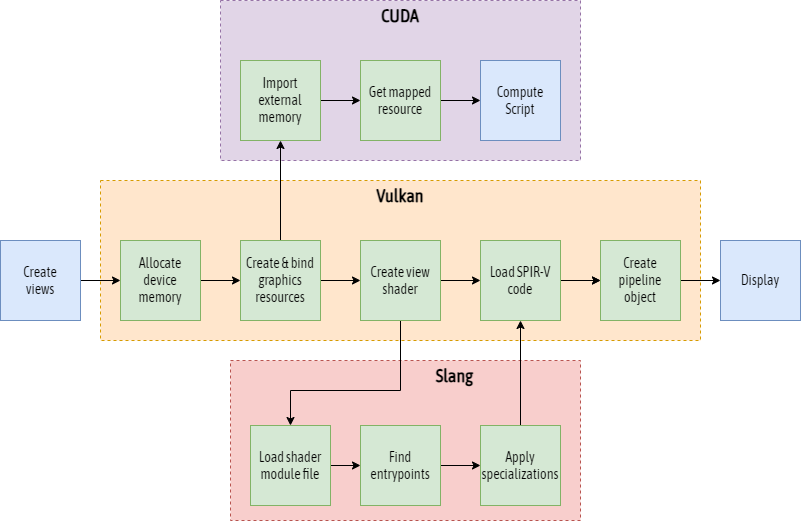}
	\caption{Overview of the view creation flow. Blue stages are handled by the library user, while green stages are managed by the library itself.}
	\label{fig:view}
\end{figure}

A Mìmir view is the set of visual features associated with a CUDA experiment composed of one or more interop data sources referenced by allocation handles. Multiple views can be created and visualized at the same time, and can reference the same allocations. Mìmir makes no assumptions about how to visualize the data sources composing a view, so a library user must provide explicitly all this information in the form of a view descriptor structure. A view description contains the view type (markers, lines or voxels), spatial domain (2D or 3D) and layout, model transformations (translation, rotation and scale) and a dictionary of various visual characteristics (position, color, size, rotation). The \texttt{createView} library function parses the supplied view description into Vulkan graphics resources needed to produce visual output, as shown in figure \ref{fig:view}. These resources are created on demand based on the view description parameters, and include buffers (vertex, index, uniform and/or storage), images, image views, samplers, descriptors, shader program and pipeline object. When the view handle is destroyed, its associated resources are cleaned up with it.

Within the library context, the view attribute descriptor contains the various visual characteristics listed above. The most important value of a view attribute is the allocation handle used as a data source, which makes the interop-mapped memory available for visualization. In addition, a view attribute must define the number and format of the values contained in the source, as allocations only reference raw device memory. The format specifier allows to describe commonly used types (\texttt{float, double, char}, etc.) and CUDA vector types (\texttt{float2}, \texttt{float3}, \texttt{float4}, \texttt{uint2}, etc...), with helper template functions for each. This scheme allows reusing allocations across different views or reference them in different attributes for the same view, with different formats for each one.

\begin{code}
\captionof{listing}{C++ Library usage example.}
\label{code:view_params}
\begin{minted}[linenos,tabsize=2,breaklines]{cpp}
using namespace mimir;
InstanceHandle instance = nullptr;
createInstance(1920, 1080, &instance); // Initialize instance

// Make interop allocation
float3 *dev_ptr = nullptr; // Pointer to interop array
AllocHandle alloc = nullptr;
allocLinear(instance, (void**)&dev_ptr, sizeof(float3)*N, &alloc);

// Describe and create view
ViewDescription desc{
    .element_count = N;
    .view_type     = ViewType::Markers;
    .domain_type   = DomainType::Domain3D;
    .properties    = PropertyDict{
        { PropertyType::Position, PropertyDescription{
            .source = alloc,
            .size   = N,
            .format = FormatDescription::make<float3>(),
        } 
    };
};
ViewHandle view = nullptr;
createView(instance, desc, &view);
\end{minted}
\end{code}

A valid view must define at least the position attribute, while the remaining attributes will use a default value when not defined. This default value (color, size, etc.) can be specified in the view descriptor or be changed at runtime with the associated function. There are also helper functions available to generate position attributes for simple cases, such as the \texttt{createGrid} function generates positions in a regular grid, and \texttt{makeImageFrame} to create a quad for showing a image texture. This is useful for experiments where these values are implicit on the studied domain, but are required for visualization.

The described procedure allows direct access to the interop data in the corresponding shader, but it is also possible to reference elements indirectly by indexing them, which can be useful for triangle meshes and value mapping. Attribute descriptions have an optional index description field for this purpose. Index descriptions are similar to attribute descriptions but only accept integral data types, so the format specifier is replaced by the size of the integer data type.

\subsection{Visualization}
The shader program code passed to the pipeline object associated to a view is internally compiled at runtime using the Slang shader library \cite{slang}. The Slang shading language follows the HLSL syntax augmented with annotations and structures, and can be compiled into a variety of targets, including other shader languages, C++ or CUDA code. The library uses the Slang compiler to emit directly the SPIR-V \cite{kessenich2018spir} intermediate representation that is passed to Vulkan code at pipeline object creation. Shader code is organized into modules that can be referenced across file through imports, and each module file can contain entry points for various shader stages, including multiple definitions for a single stage. This is used in the library to organize modules in separate files for the various view types and properties, and select entry points based on parameters such as the spatial domain type. When creating a view, only modules relevant to that view are loaded and linked at shader compilation, so that even code in different files can be made visible when needed.

A feature of Slang shaders is the possibility of defining interfaces, with their required methods, and conforming structures that implement said methods. When a shader function (including entrypoints) uses an interface type as parameter, it is possible to select a conforming implementation at shader compile time by passing its typename as a specialization. Within the library, specializations are used to define behavior of view properties inside the shader according to the resources referenced in the description for those properties, which include buffer and texture bindings. The format description inside a property allows the shader to adjust to the defined format, so it is possible for example to represent a property buffer as an array of \texttt{float2}, \texttt{double3} and other CUDA vector type variants, including functions to read those types and convert them into the expected format for that property (e.g. \texttt{float4} for positions and colors in shader code). Views with a missing property description use a basic specialization that does not bind any resource and reads its value from uniform memory, which corresponds to the default value set in the view structure.

Slang capabilities can also be used to choose how to draw elements of a visualization, providing further customization of the visual results through the view description structure. As an example, antialiased 2D markers \cite{Rougier2014Antialiased2D} were ported from the GLSL reference code to Slang implementations, with specializations for shape (disc, diamond, arrow, etc.) and style (filled, stroked or outlined). Selection of the marker shape and type is available through the view descriptor in the \texttt{options} parameter, implemented as a variant (type-safe C++ union types). This allows having separate option fields for each visualization type without the need of having all fields in a single uber-structure or use inheritance.

\subsection{Synchronization}

Upon calling the display function, the library spawns a separate thread running the rendering loop to avoid interference with the compute workflow. CUDA and Vulkan device calls associated with the compute-rendering loop run in mutual exclusion, where the critical section corresponds to the CUDA kernel(s) writing on the interop memory. A lack of synchronization does not compromise the physical computations, but may lead to visual artifacts as the rendered frame can contain information from older and newer states of the simulation, compromising the quality for visual inspection. Another argument for synchronization arises from the need to regulate access to GPU computing power between both components, so that the rendering-presentation loop does not saturate computing performance by showing superfluous frames of an already presented state.

\begin{figure}[H] \centering
	\includegraphics[scale=0.5]{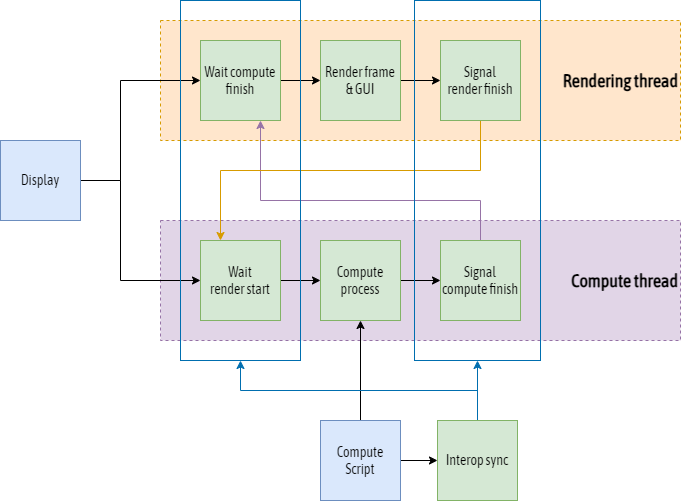}
	\caption{Overview of the interop synchronization scheme. Blue stages are initiated by the user, while green stages are managed by the library.}
	\label{fig:sync}
\end{figure}

The library offers two modes for controlling synchronization behavior through separate function calls when initiating display. The \texttt{display} method is the most simple case and executes a critical compute section for a number of iterations, handling synchronization internally. Alternatively, the \texttt{displayAsync} function starts display but does not perform any additional work, instead requiring to explicitly surround critical compute code with the \texttt{prepareViews} and \texttt{updateViews} functions. This enables more advanced synchronization patterns, such as reading from one buffer and writing into another, all in parallel. This \say{ping-pong} scheme allows displaying any of the two buffers according to the simulation. Examples of simulations that work this way are cellular automata or even then n-body simulation where two copies of the particles are kept alternating on which is the read / write buffer. Listing \ref{code:sync} shows the rendering loop using this scheme. In these cases, both views should be created with the same description excepting the source allocation for all relevant properties, and only one view at a time should have visibility turned on at the same time. If other parameters are different across views, the visual continuity between swaps would be lost, and if both views are visible at the same time, the display would show artifacts due to z-fighting.

\begin{code}
\captionof{listing}{Alternating buffer view example.}
\label{code:sync}
\begin{minted}[linenos,tabsize=2,breaklines]{cpp}
displayAsync(instance); // Display with explicit sync management
for(int i=0; i < NUMBER_STEPS; ++i){
    prepareViews(instance); // Start of compute section

    // Update data in d_write from previous step in d_read
    compute_kernel<<<grid, block>>>(d_write, d_read);
    v1->toggleVisibility();
    v2->toggleVisibility();
    
    updateViews(instance); // End of compute section
}
\end{minted}
\end{code}

\section{Experimental Results}
\label{sec:results}

Mìmir is implemented as a \texttt{C++} library using CUDA as the compute layer, Vulkan as its graphics layer and interop between the two to perform efficient real-time visualization of physics-based experiments. The code uses the CMake build system to compile and install the library target, and includes sample code for generating visualizations for various view types, use cases and experiments, including the ones presented in the following sections.

\subsection{Performance Benchmark}
\begin{figure}[H] \centering
	\subcaptionbox{$N=10^3$.}{\includegraphics[scale=0.09]{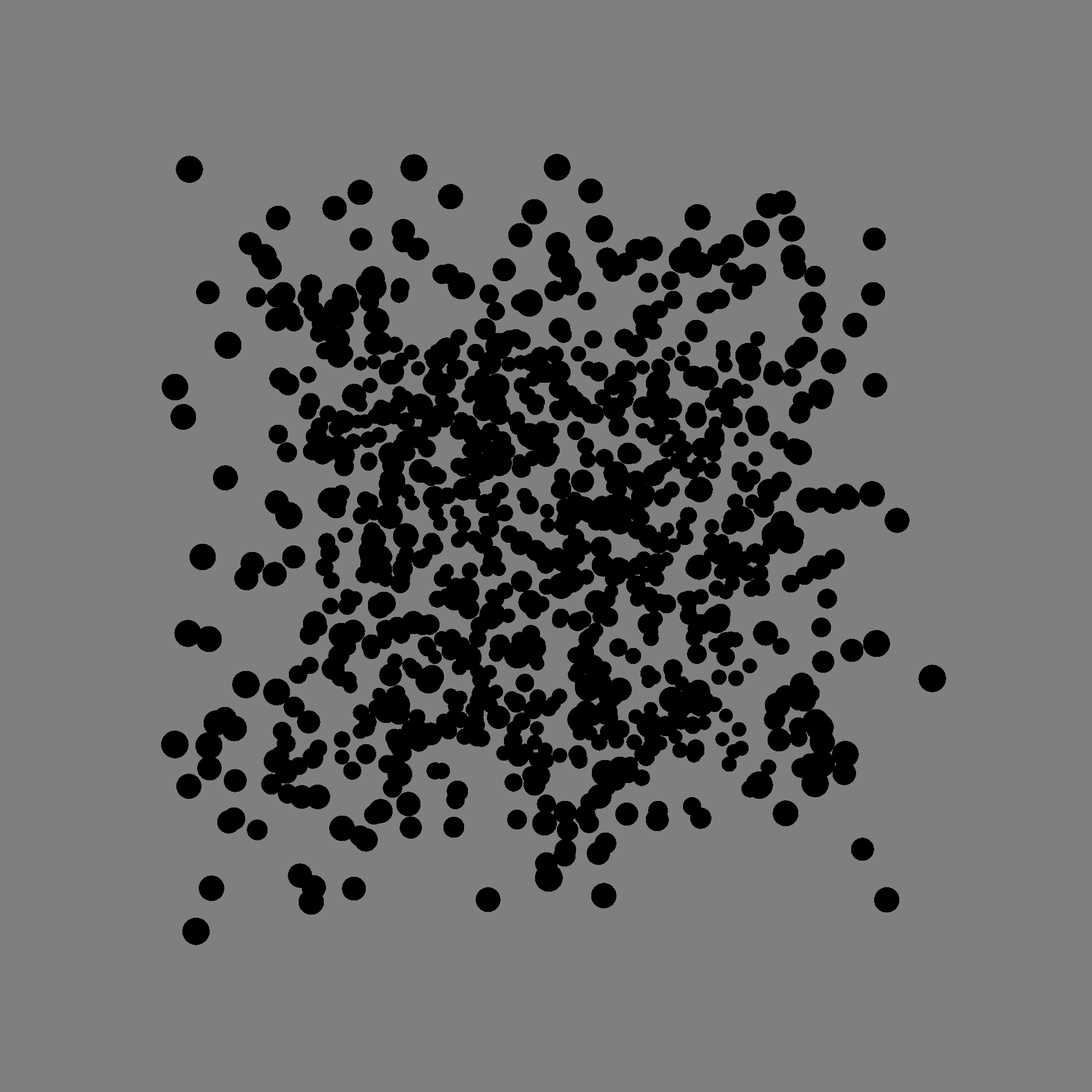}}
	\hfill
	\subcaptionbox{$N=10^4$.}{\includegraphics[scale=0.09]{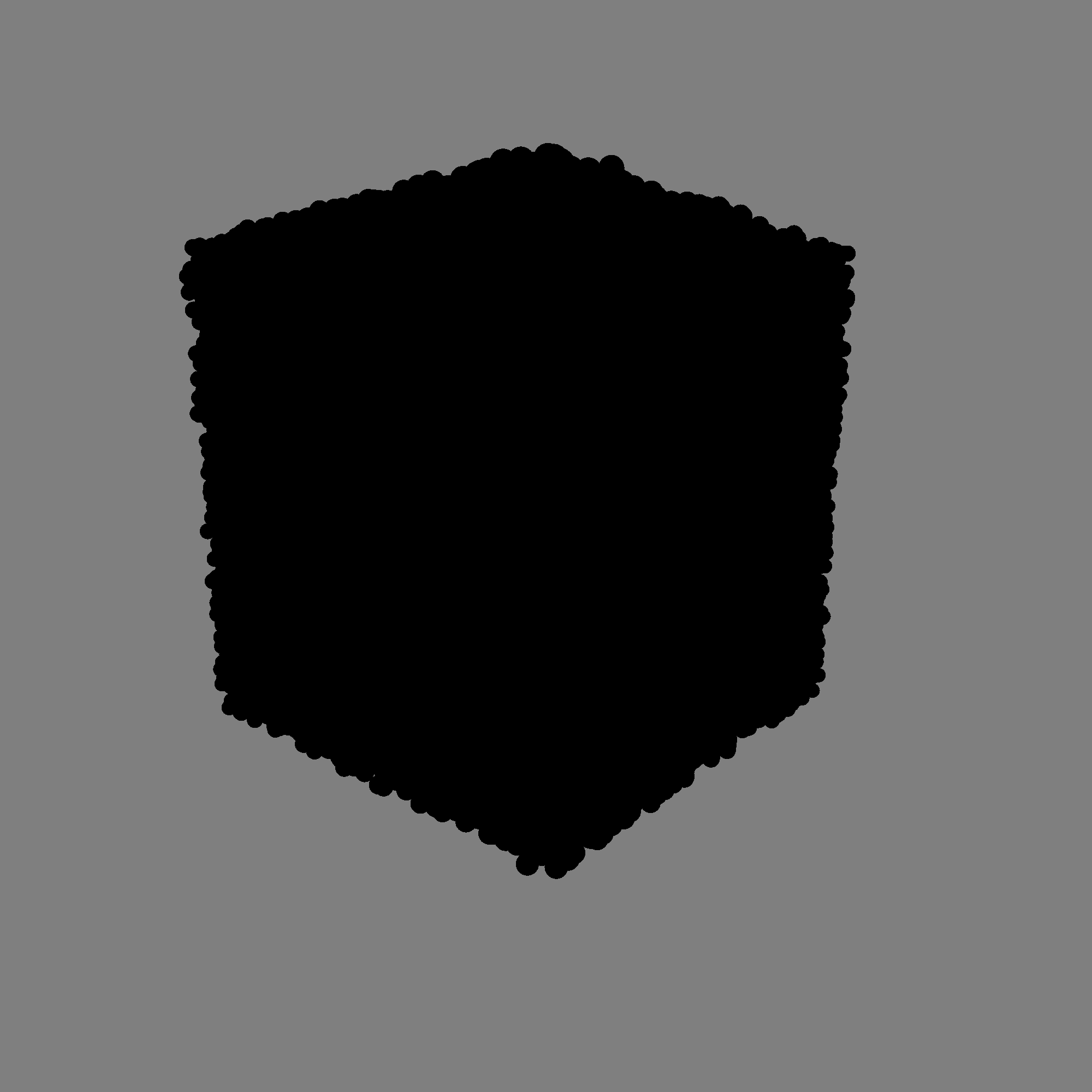}}
	\caption{Point cloud visualization experiment visualized with Mìmir.}
	\label{fig:benchmark-random-particles}
\end{figure}

The benchmark consists in visualizing a 3D point cloud with randomly generated starting positions and Brownian motion, with no additional physics involved (other physical models are presented in sub Section \ref{subsec:sample-applications}). Points are rendered as front-facing disc markers. All points are rendered with the same size and color (see Figure \ref{fig:benchmark-random-particles}). The setup stage initializes all $n$ points with random positions using CUDA pseudo random generator states in parallel. This process occurs just once at the beginning, therefore its running time is not included in the measurement. Then, at each timestep, the simulation executes a CUDA kernel for moving the points with Brownian motion, which updates the visualization handled by Mìmir in real-time, because it is rendering the same data. The benchmark visualization is generated using the code listing \ref{code:view_params}, for various input sizes and display configurations. The results were obtained using a desktop platform with the following specs and compilation settings:
\begin{itemize}
    \item \textit{CPU:} AMD Ryzen 7 3700X CPU / 16GB RAM
    \item \textit{GPU:} GeForce RTX 2070 SUPER (8GB VRAM)
    \item \textit{OS: Ubuntu 22.04 LTS}
    \item \textit{C++:} \texttt{gcc 11.3}, \texttt{C++17}
    \item \textit{CUDA:} Version \texttt{12.2}, driver version \texttt{535.54.03}
    \item \textit{Vulkan:} Version \texttt{1.3.236}
    \item \textit{Display:} size 24''; resolution $1920 \times 1080$; refresh rate 144 FPS
\end{itemize}

\begin{figure}[H] \centering
    \includegraphics[width=\linewidth]{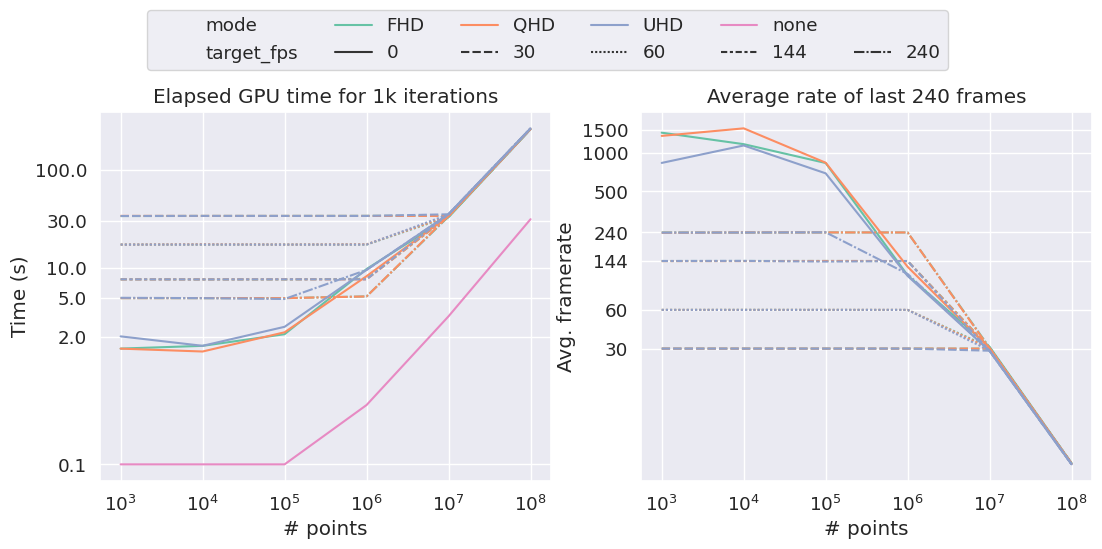}
	\caption{Performance results for various combinations of framebuffer size and target FPS.}
	\label{plot:performance}
\end{figure}

Figure \ref{plot:performance} shows performance results for different combinations of configurations, where FPS for the renderless base case is zero. Frame rate limits are shown to be correctly applied when set, and FPS degradation starts occurring starting at $N = 10^6$ for the largest FPS target. This is consistent with the total GPU time (rendering and compute), which also grows sharply starting from that value. For $N = 10^9$ rendering performance is no longer responsive and acquiring a single frame for rendering takes seconds. Elapsed time increases with lower FPS targets compared to the renderless case because compute is being stalled in order to match the rendering pace being forcibly delayed.

\begin{figure}[H] \centering
	\includegraphics[width=\linewidth]{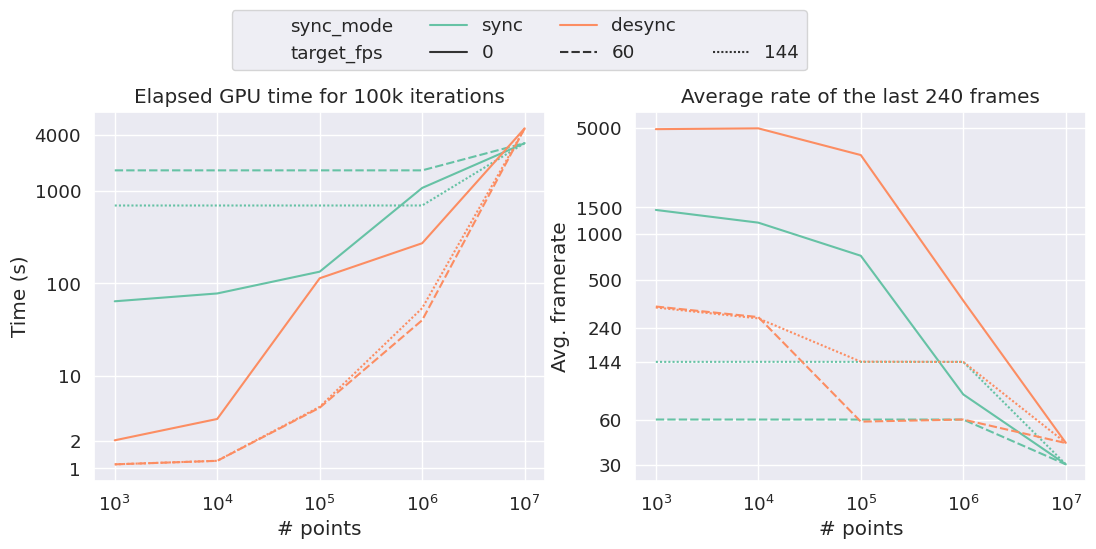}
	\caption{Performance results for different combinations of synchronization mode and target FPS.}
	\label{plot:desync_performance}
\end{figure}

Figure \ref{plot:desync_performance} shows the difference in performance when synchronization is disabled. This means compute and rendering workflows run at full speed without any Vulkan semaphores to wait for each other to finish at each frame. Visualization is fast for small point clouds as expected, though performance decreases with input size until it becomes worse than synchronized display. By disabling synchronization for demanding visualizations, CUDA and Vulkan workflows end up competing with each other for GPU usage. For this reason, it is important to note that interop synchronization has the key property of ensuring GPU processing power is properly shared between compute and rendering tasks, which may be even more important than the original intent of preventing visual artifacts from simultaneous GPU memory access.

\begin{figure}[H] \centering
    \includegraphics[width=\linewidth]{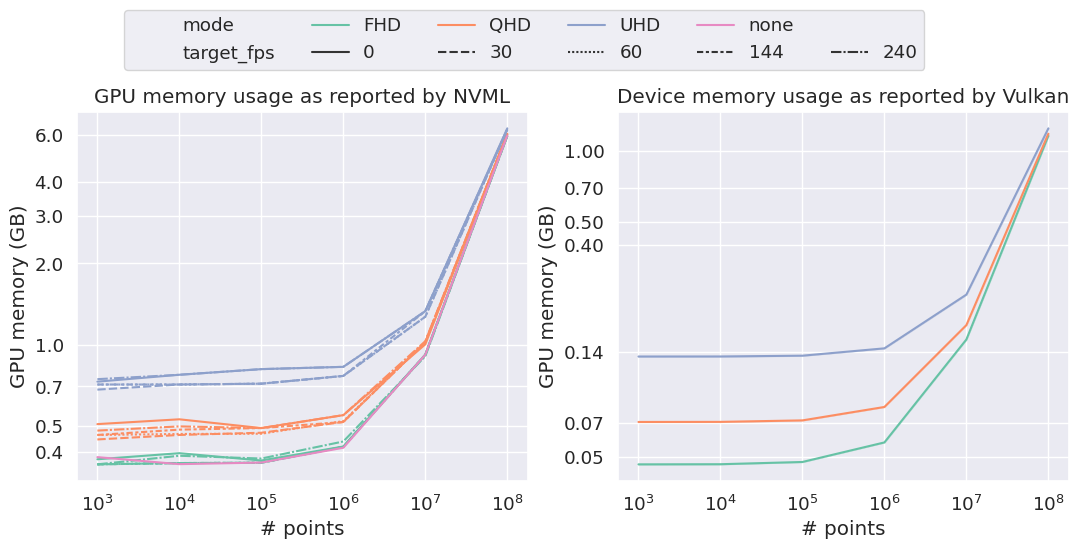}
	\caption{Device memory usage across different experiment runs, measured from CUDA and Vulkan APIs.}
	\label{plot:usedmem}
\end{figure}

Figure \ref{plot:usedmem} shows device memory usage from Vulkan context and from the whole GPU. As expected, Vulkan memory use does not depend on target FPS, but it does depend on framebuffer size and input size, because interop memory allocation is handled by the Vulkan API. Base memory usage in Vulkan is zero because allocation in that case is via \texttt{cudaMalloc}, but it gets correctly measured through NVML. It is worth noting that NVML does measure differences in memory usage between FPS targets, and the ordering is as expected. That is, base uses the least memory, then the resolutions starting from the lowest, and the resolutions themselves are ordered by the target FPS from least to most (target $0$ is unlimited FPS). Memory usage is still different across FPS targets even at $N = 10^7$, where figure \ref{plot:performance} shows all cases are displaying at approximately 24 FPS. Differences in Vulkan memory usage remain constant across the three resolutions; UHD memory is roughly double that of QHD, which in turn is about double that of FHD. This is consistent with memory usage depending majorly on framebuffer size.

\begin{figure}[H] \centering
    \includegraphics[width=\linewidth]{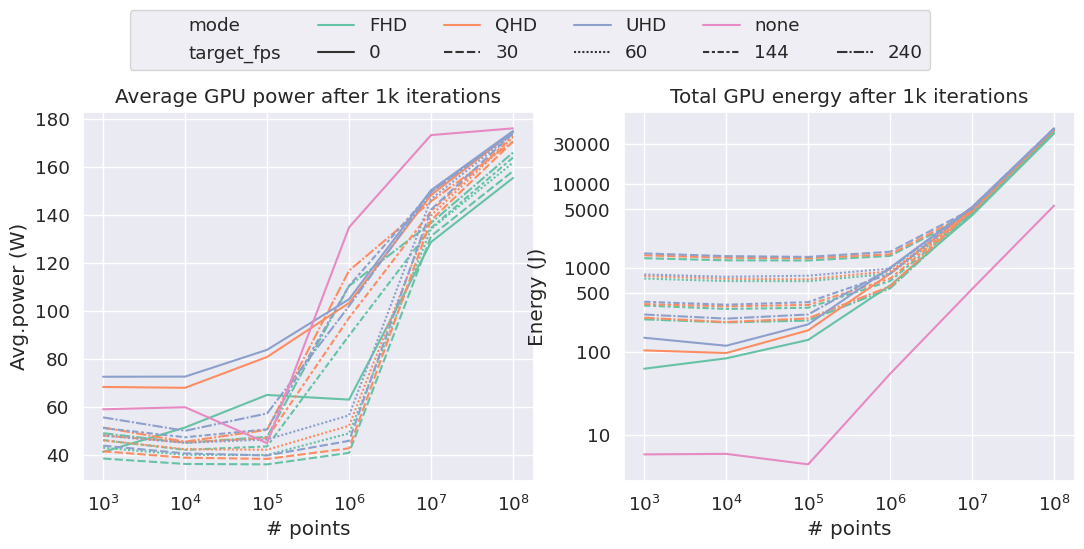}
	\caption{Device energy metrics for concurrent compute and graphics work.}
	\label{plot:power}
\end{figure}

Figure \ref{plot:power} shows GPU average power after all iterations. As expected, total energy is the lowest without display as there is no Vulkan work being done. Average power in this mode is the highest for larger input sizes, which is consistent with the lower time needed for completing the execution. Higher resolution and higher FPS targets contribute to greater power usage (here FPS target zero means the highest possible FPS value), but all these quantities are dominated by the simulation size when it is large enough.

\begin{figure}[H] \centering
	\includegraphics[width=\linewidth]{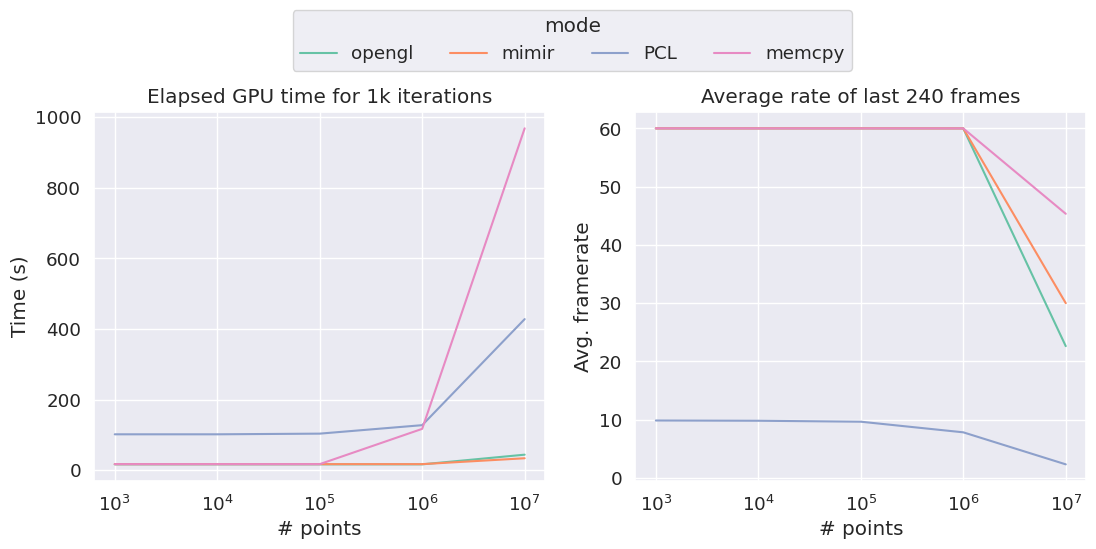}
	\caption{Performance results for various input sizes and execution modes.}
	\label{plot:alt-perf}
\end{figure}

For comparison, the benchmark visualization is implemented using various backends with equivalent visual output, whose performance is shown in figure \ref{plot:alt-perf}. One approach is to display the moving point cloud using the visualization capabilities provided by the point cloud library (PCL) \cite{Rusu_ICRA2011_PCL}. Although PCL does contain CUDA versions for most cloud processing methods, it does not provide low level GPU visualization capabilities and defers to VTK for rendering.

Another alternative is to display the point cloud with OpenGL, getting comparable performance with a slight advantage to the Vulkan backend, matching previous comparison results between the two platforms \cite{hong2016parallel}. In this case, the difference can be attributed to the implicit host-device and device-device synchronization in most OpenGL calls, compared to Vulkan where the task is up to the program specifics. As more Vulkan-only optimizations are implemented in the library, it should be possible to widen this gap. However, it is also possible that newer OpenGL extensions may keep it competitive in this regard.

It is also possible to use Mìmir for display but performing the compute workload in CPU using OpenMP and PCG random number generator \cite{oneill:pcg2014} for efficiency, and transferring the states to interop memory at each iteration. In Mìmir terms, the compute critical section used in this case is only the \texttt{cudaMemcpy} call updating the position data for the previous iteration. 

\subsection{Sample applications}
\label{subsec:sample-applications}
In this Section we present real use cases, corresponding to computational physics simulations with existing CUDA implementations, now modified to include Mìmir for real-time visualization. 

\subsubsection{Colloidal simulation}
This visualization shows a system of self-assembling colloidal molecules \cite{PhysRevE.91.052304} in 2D unstructured space. This research explored the feasibility of using 2D Delaunay triangulations to solve particle overlaps at each time step \cite{CARTER2018148}, using a GPU parallel edge-flip algorithm that updates and repairs the triangulation after each simulation step. This simulation is able to detect invalid scenarios before applying the simulation step, such as particles passing through others. In this simulation, there is the main visualization which are the particles themselves moving, and alternatively one could need to visualize the underlying triangulation that handles the neighbor interactions. The Mìmir library is capable of visualizing both in real-time.

\begin{figure}[ht!] \centering
	\subcaptionbox{Initial system state.}{\includegraphics[scale=0.09]{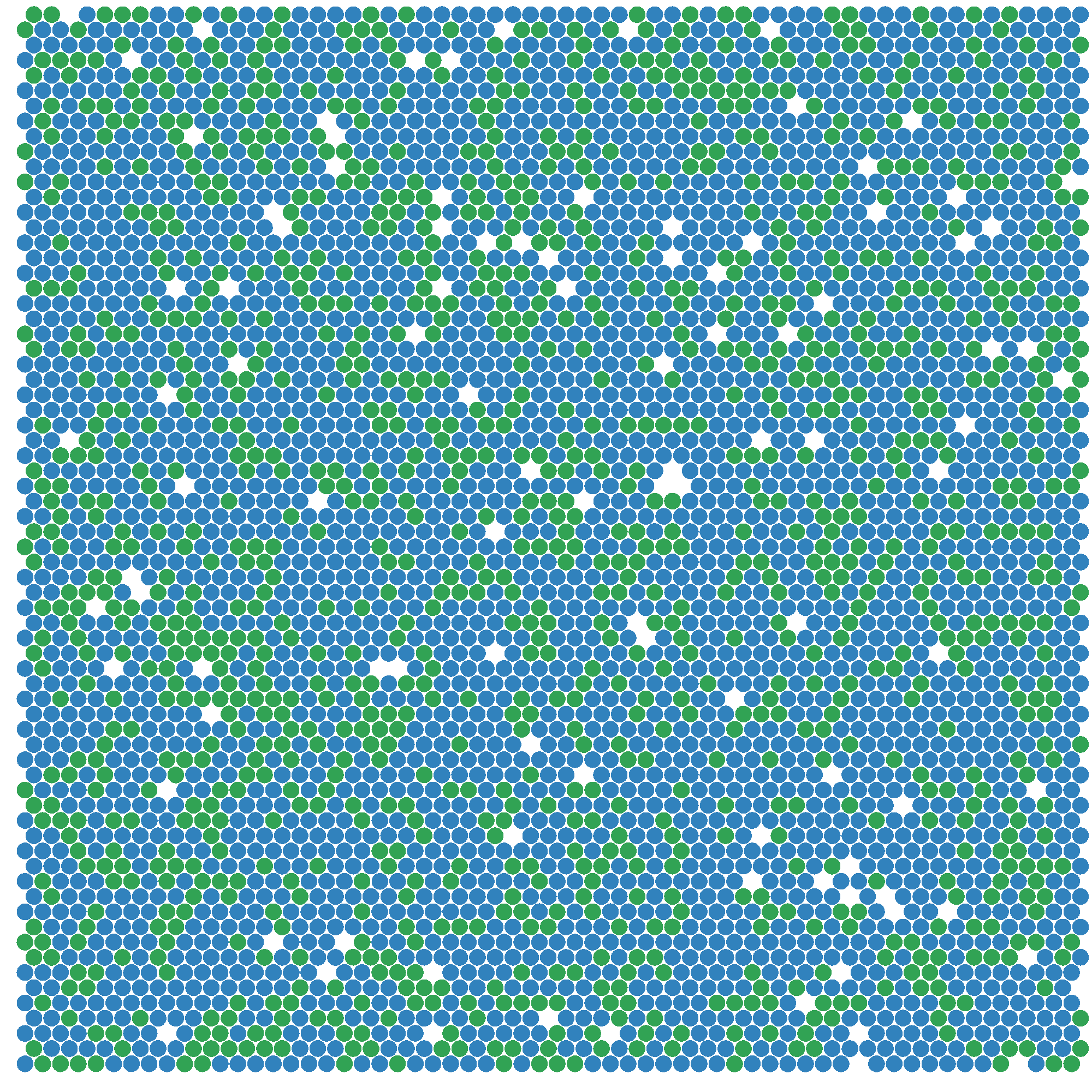}}
	\hfill
	\subcaptionbox{Delaunay triangulation overlaid with particle positions.}{\includegraphics[scale=0.09]{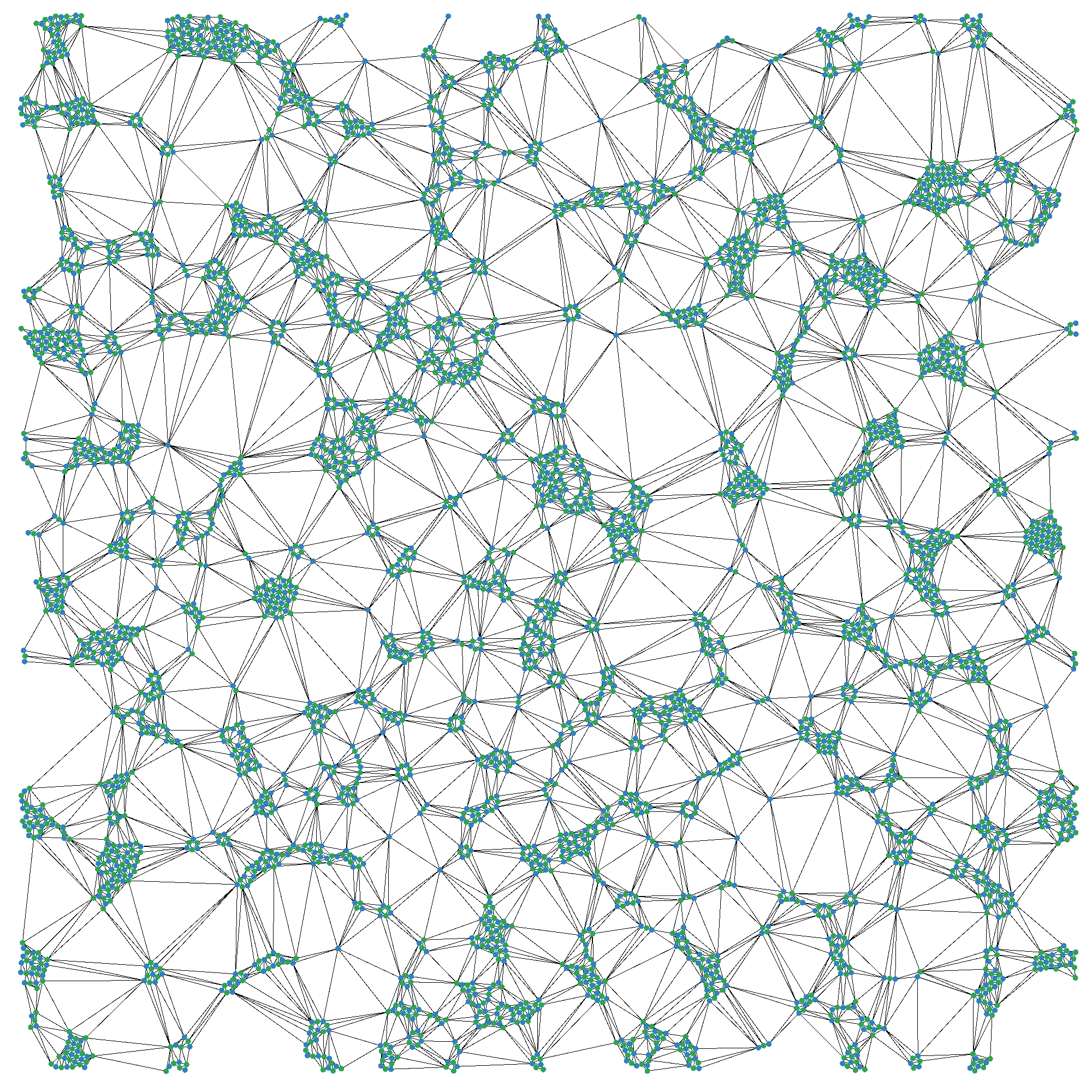}}
	\caption{Visualization of colloidal particle system using Mìmir both as particles and geometry mesh.}
	\label{fig:colloids0}
\end{figure}

\begin{figure}[ht!] \centering
	\subcaptionbox{Configuration 1.}{\includegraphics[scale=0.09]{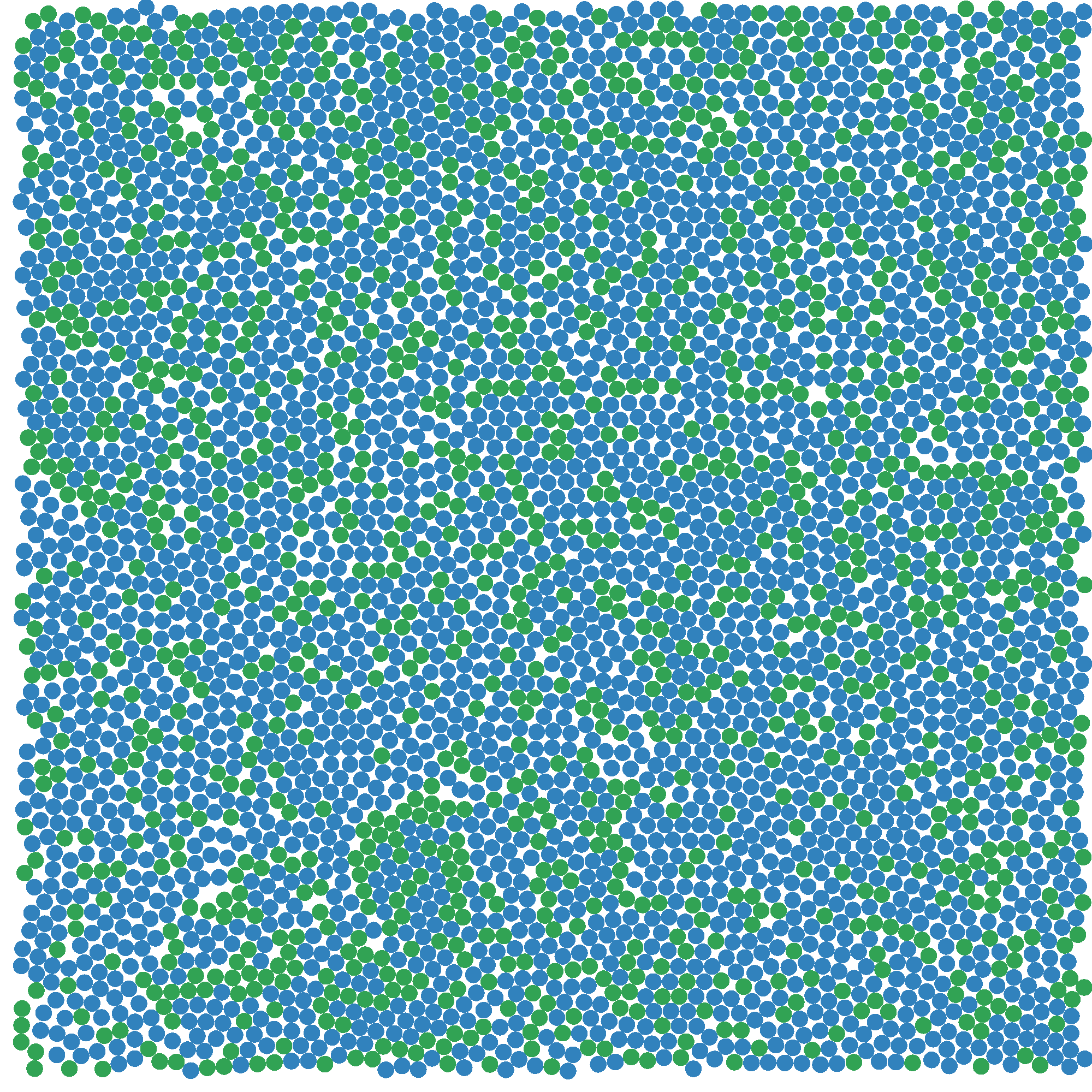}}
	\hfill
	\subcaptionbox{Configuration 2.}{\includegraphics[scale=0.09]{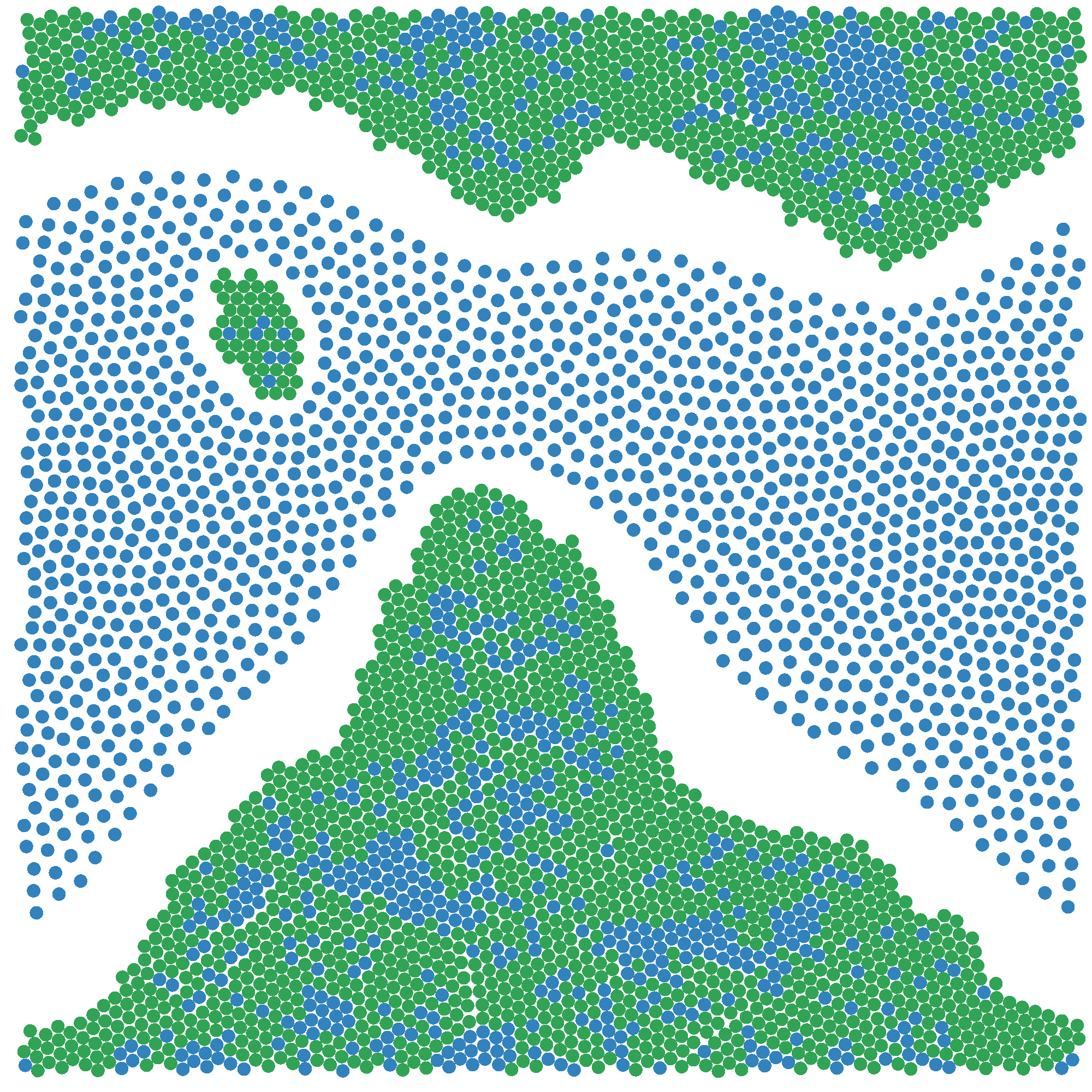}}

	\subcaptionbox{Configuration 3.}{\includegraphics[scale=0.09]{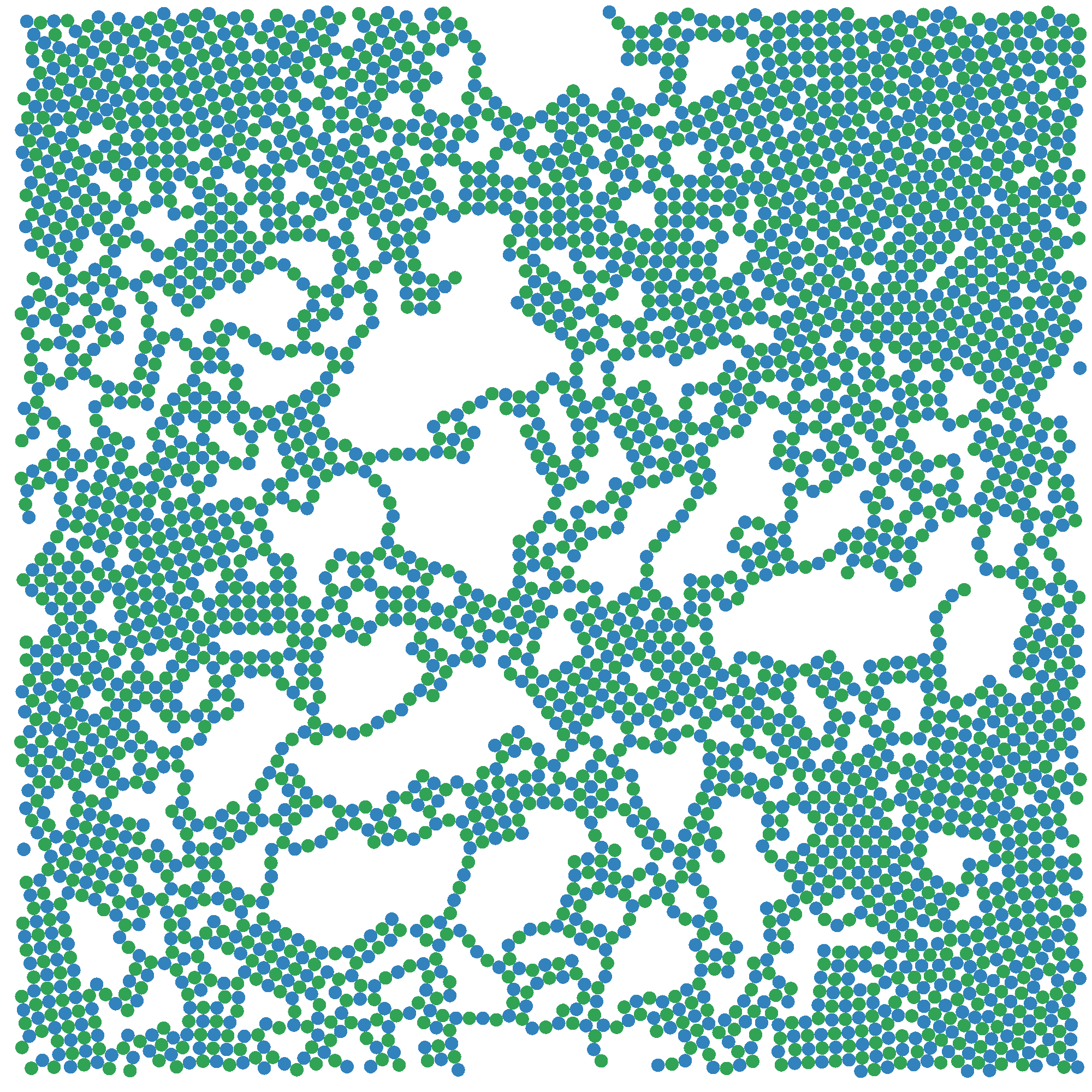}}
	\hfill
	\subcaptionbox{Configuration 4.}{\includegraphics[scale=0.09]{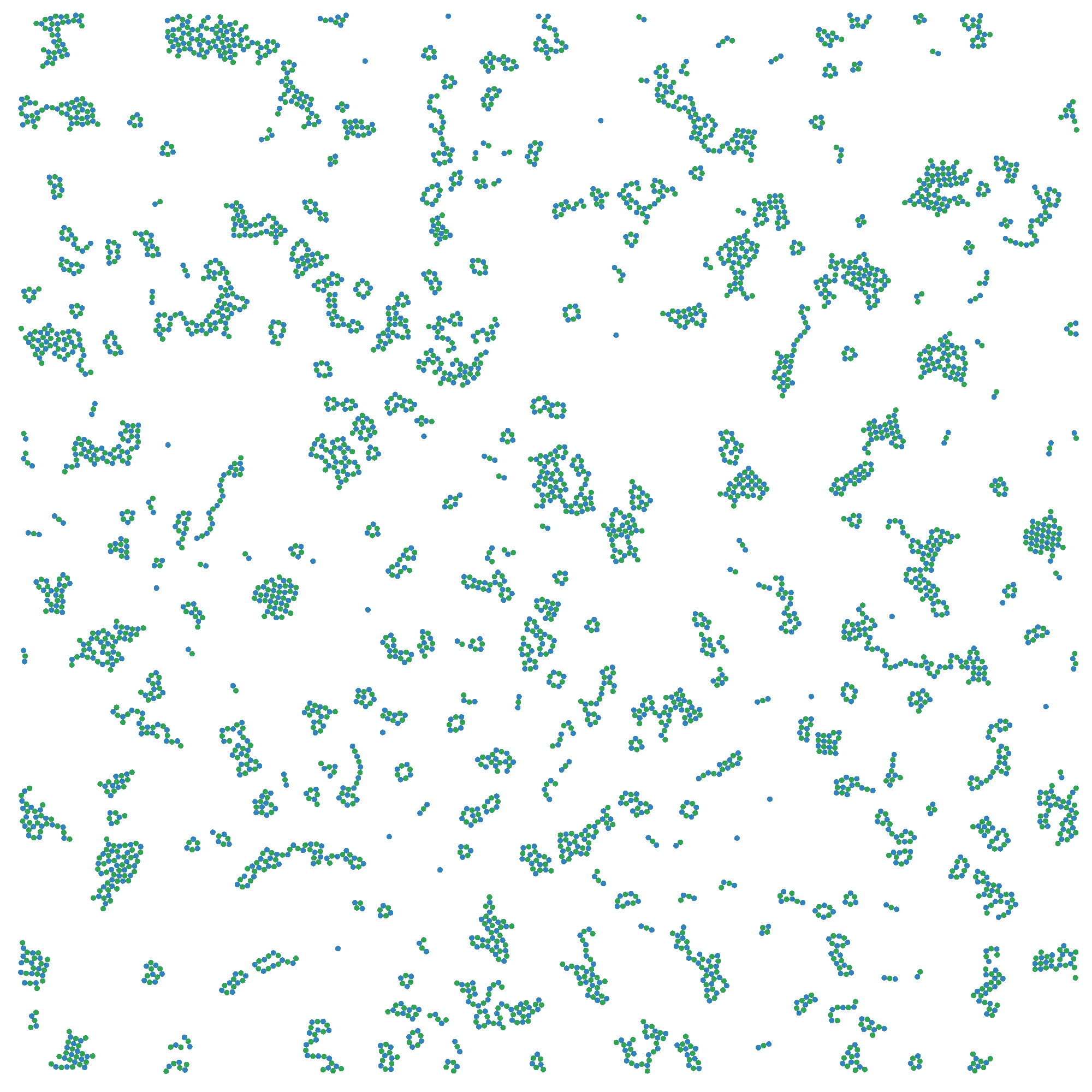}}
	\caption{Visualization of colloidal particle system using Mìmir as particles.}
	\label{fig:colloids1}
\end{figure}

The visualization shows molecular colloid systems for various configurations of densities, concentrations and particle types, where each type $i$ is characterized by its parameters $(\alpha_i, \mu_i)$. The initial state is generated by reservoir sampling of uniformly distributed particles to generate configurations of the desired particle density and concentration rates per type. A 2D periodical Delaunay triangulation is generated in host memory using particle positions as vertices, which then is fully maintained in GPU memory by the parallel edge-flip executed after particles move after integration. Figure \ref{fig:colloids0} shows visualizations with Mìmir for the initial configuration and the Delaunay triangulation with the current particle positions overlaid on top. Figure \ref{fig:colloids1} shows more screenshots of the real-time visualization of colloids as free floating particles.

It is worth mentioning that before the existence of Mìmir, visualizations on this work had to be generated using a script in the Processing sketchbook software \footnote{\url{https://processing.org/}}, where it was necessary to copy the entire simulation state to host RAM memory at each time step, save it to a file, read the file in the Processing script and save the resulting image back to a file, and finally use the ImageMagick software \cite{imagemagick} to generate the image sequence. With Mìmir, the inclusion of visualization allows the programmer to keep the attention on the simulation process as the visualization is immediate on the simulation data. 

The visualization setup code shown in listing \ref{code:colloids} uses a double buffering scheme described in code \ref{code:sync} to update particle positions and edge endpoints at the end of each iteration, with alternating read/write roles for each buffer. This simulation approach is also known as \textit{ping pong}. For this visualization it is necessary to replace \texttt{cudaMalloc} allocations of position, edge and triangle index arrays with the library interop allocations. Particle and edge allocations are mapped to the ping-pong arrays and assigned to their respective views. Particle colors are accessed indirectly through particle type indices stored in the \texttt{view\_types} allocation. Similarly, edge views are created using indirect access for positions, where particle coordinates are the primary source and triangle vertex indices are used to access it. Views created for the double-buffering scheme use the same parameters except for the source handle passed to the position property. This is to ensure there is no visual difference between the two buffers. Consequently, only one view remains active (controlled by the \texttt{visible} parameter and the \texttt{toggleViews} method).

\begin{code}
\captionof{listing}{N-Body colloids setup code.}
\label{code:colloids}
\begin{minted}[linenos,tabsize=2,breaklines]{cpp}
// Particle positions
allocLinear(instance, (void**)&devicedata_.positions[current_read], pos_bytes, &view_pos[current_read]);
allocLinear(instance, (void**)&devicedata_.positions[current_write], pos_bytes, &view_pos[current_write]);

// Particle types / colors
allocLinear(instance, (void**)&devicedata_.types, sizeof(int) * params_.num_elements, &view_types);
allocLinear(instance, (void**)&devicedata_.colors, sizeof(float4) * NUM_TYPES, &view_colors);

ViewDescription vp;
vp.element_count = params_.num_elements;
vp.extent        = {l, l, 1};
vp.domain_type   = DomainType::Domain2D;
vp.view_type     = ViewType::Markers;
vp.properties[PropertyType::Position] =
{
    .source = interop[current_read],
    .size   = params_.num_elements,
    .format = FormatDescription::make<double2>(),
};
vp.properties[PropertyType::Color] =
{
    .source  = view_colors,
    .size    = NUM_TYPES,
    .format  = FormatDescription::make<float4>(),
    .indices = view_types,
    .index_size = sizeof(int),
};
createView(instance, &vp, &particle_views[current_read]);
particle_views[current_read]->default_size = vp.extent.x / l;

vp.properties[PropertyType::Position].source = interop[current_write];
createView(instance, &vp, &particle_views[current_write]);
particle_views[current_write]->default_size = vp.extent.x / l;
particle_views[current_write]->visible = false;

// Edges
allocLinear(instance, (void**)&devicedata_.triangles, sizeof(int3) * delaunay_.num_triangles, &view_tri);

ViewDescription vpe;
vpe.element_count = delaunay_.num_triangles * 3;
vpe.extent        = {l, l, 1};
vpe.domain_type   = DomainType::Domain2D;
vpe.view_type     = ViewType::Edges;
vpe.properties[PropertyType::Position] =
{
    .source  = interop[current_read],
    .size    = params_.num_elements,
    .format  = FormatDescription::make<double2>(),
    .indices = view_tri,
    .index_size = sizeof(uint),
};
createView(instance, &vpe, &edge_views[current_read]);

vpe.properties[PropertyType::Position].source = interop[current_write];
createView(instance, &vpe, &edge_views[current_write]);
edge_views[current_write]->visible = false;
\end{minted}
\end{code}

Listing \ref{code:colloids2} shows the update step executed at each iteration. The critical interop section is declared between \texttt{prepareViews} and \texttt{updateViews}, which includes the integration step, overlap correction, Delaunay edge flip and ensuring system validity. At the end of \texttt{correctOverlaps}, the values of \texttt{current\_read} and \texttt{current\_write} are swapped, and the displayed view should change accordingly. The \texttt{toggleVisibility} calls for particles and edges are used to switch the displayed view: the currently visible view becomes hidden and vice versa.

\begin{code}
\captionof{listing}{N-Body colloids display loop.}
\label{code:colloids2}
\begin{minted}[linenos,tabsize=2,breaklines]{cpp}
prepareViews(instance);

integrateShuffle();
updateTriangles();
checkTriangulation();
updateDelaunay();
correctOverlaps();

particle_views[current_read]->toggleVisibility();
particle_views[current_write]->toggleVisibility();

edge_views[current_read]->toggleVisibility();
edge_views[current_write]->toggleVisibility();

updateViews(instance);
\end{minted}
\end{code}



\subsubsection{Potts model}

This sample visualizes a two-dimensional q-state GPU Potts model \cite{FERRERO20121578} in 2D structured space. The implementation used in this sample is available online\footnote{Github source at \url{https://github.com/ezeferrero/Potts-Model-on-GPUs}}. The simulated $L\times L$ lattice spin system is partitioned into halved sets of $\frac{L^2}{2}$ \say{black} and $\frac{L^2}{2}$ \say{white} cells, updated separately in parallel using a checkerboard scheme. Cells in the black and white sets are stored continuously in their respective arrays.

\begin{figure}[ht!] \centering
	\subcaptionbox{Original layout of black and white cell sets.}{\frame{\includegraphics[scale=0.09]{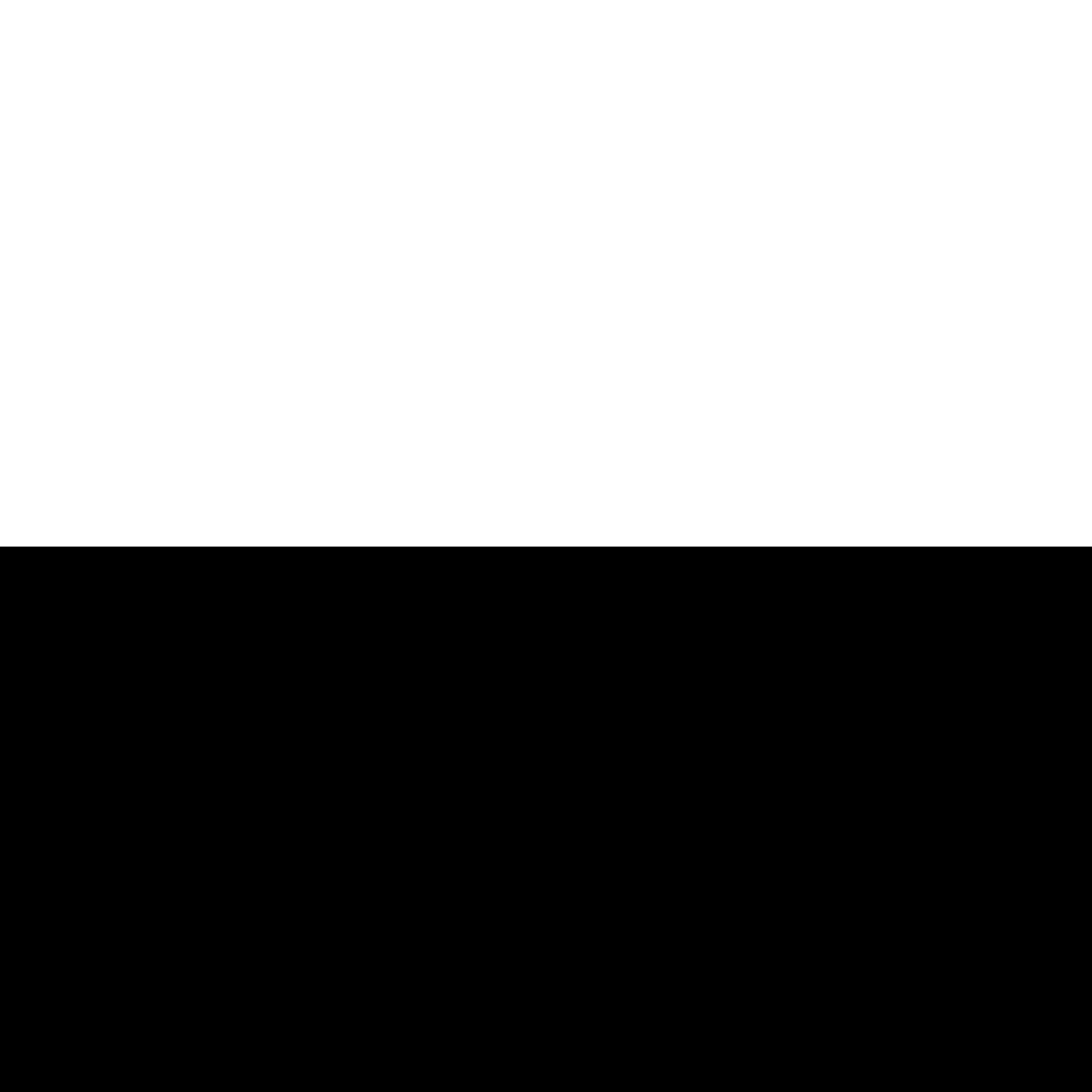}}}
	\hfill
	\subcaptionbox{Closeup of lattice arranged in a checkerboard pattern.}{\frame{\includegraphics[scale=0.09]{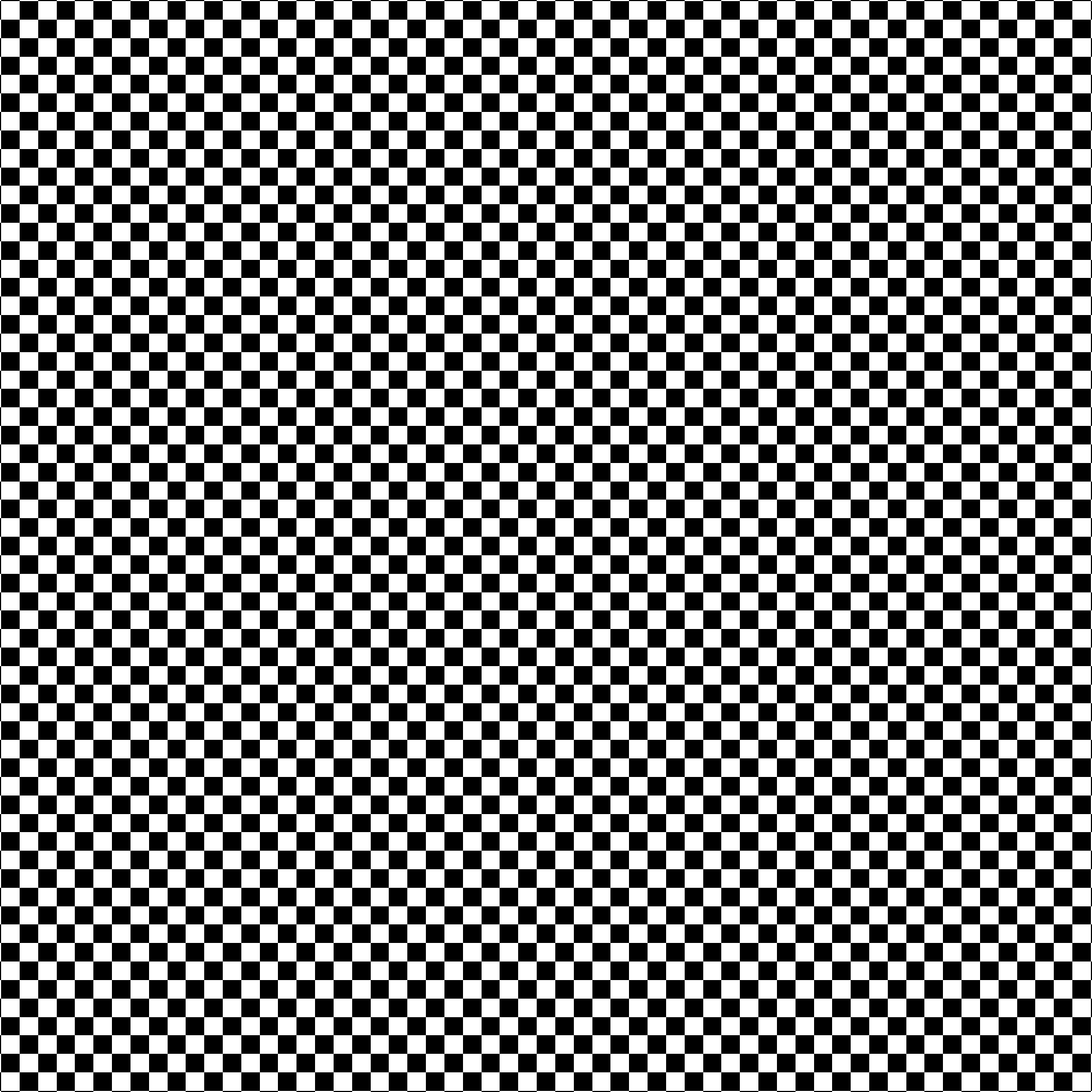}}}
    \caption{Visualization of assembled lattice layout using Mìmir.}
	\label{fig:potts0}
\end{figure}

The lattice is visualized using a single view object, whose state is updated from the contents of the black and white packed arrays, using a custom kernel \texttt{writeGrid} to perform this operation after each timestep. Figure \ref{fig:potts0} shows synthetic initial configurations visualized with Mìmir, where cells from the black and white sets are assigned initial states assigned to matching colors. This physics simulation serves as an example of how the library can be used to perform intermediate visual checks, in this case showing that \texttt{writeGrid} kernel writes the lattice data in the expected layout.

\begin{figure}[ht!] \centering
	\subcaptionbox{$T=0.71$.}{\includegraphics[scale=0.09]{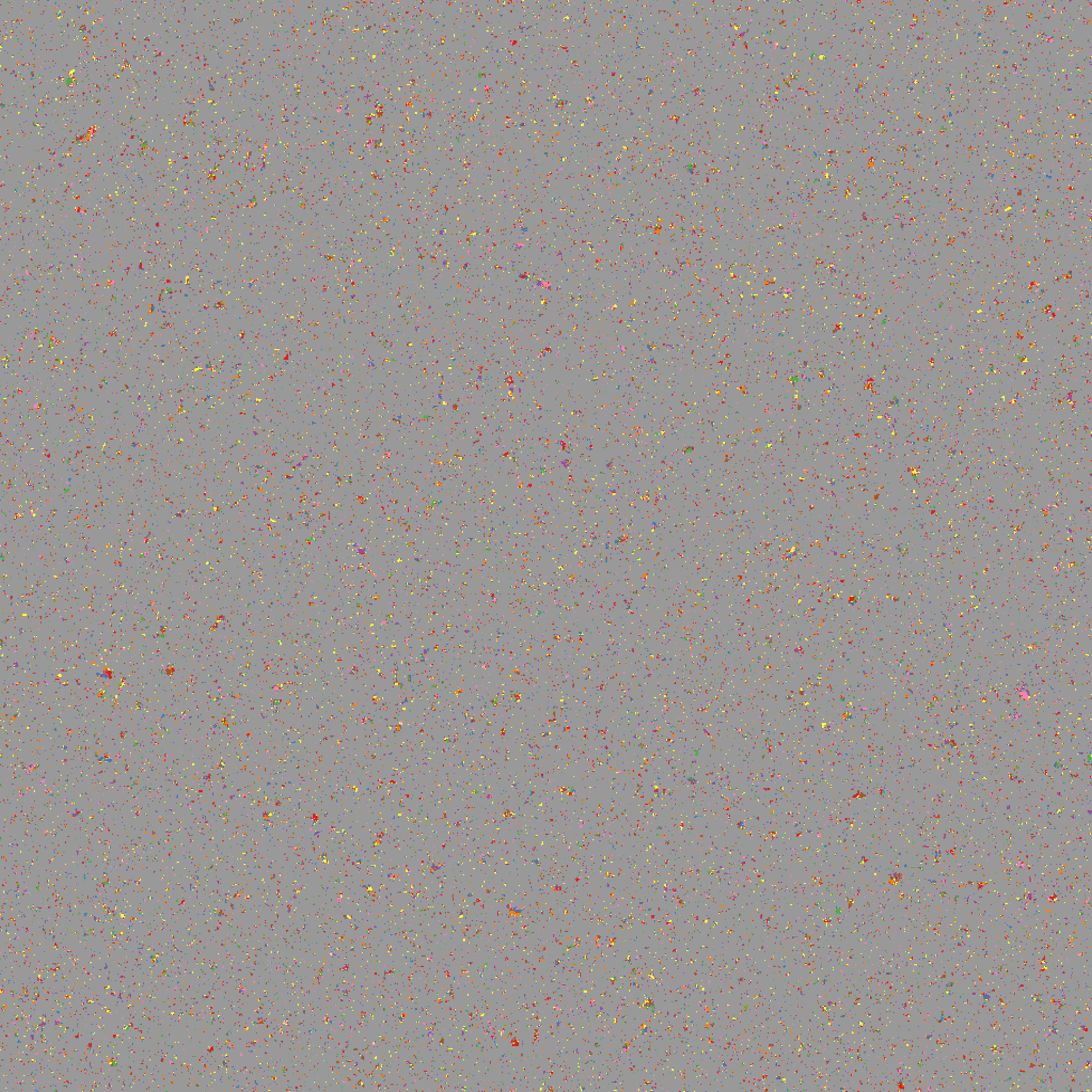}}
	\hfill
	\subcaptionbox{$T=0.722$.}{\includegraphics[scale=0.09]{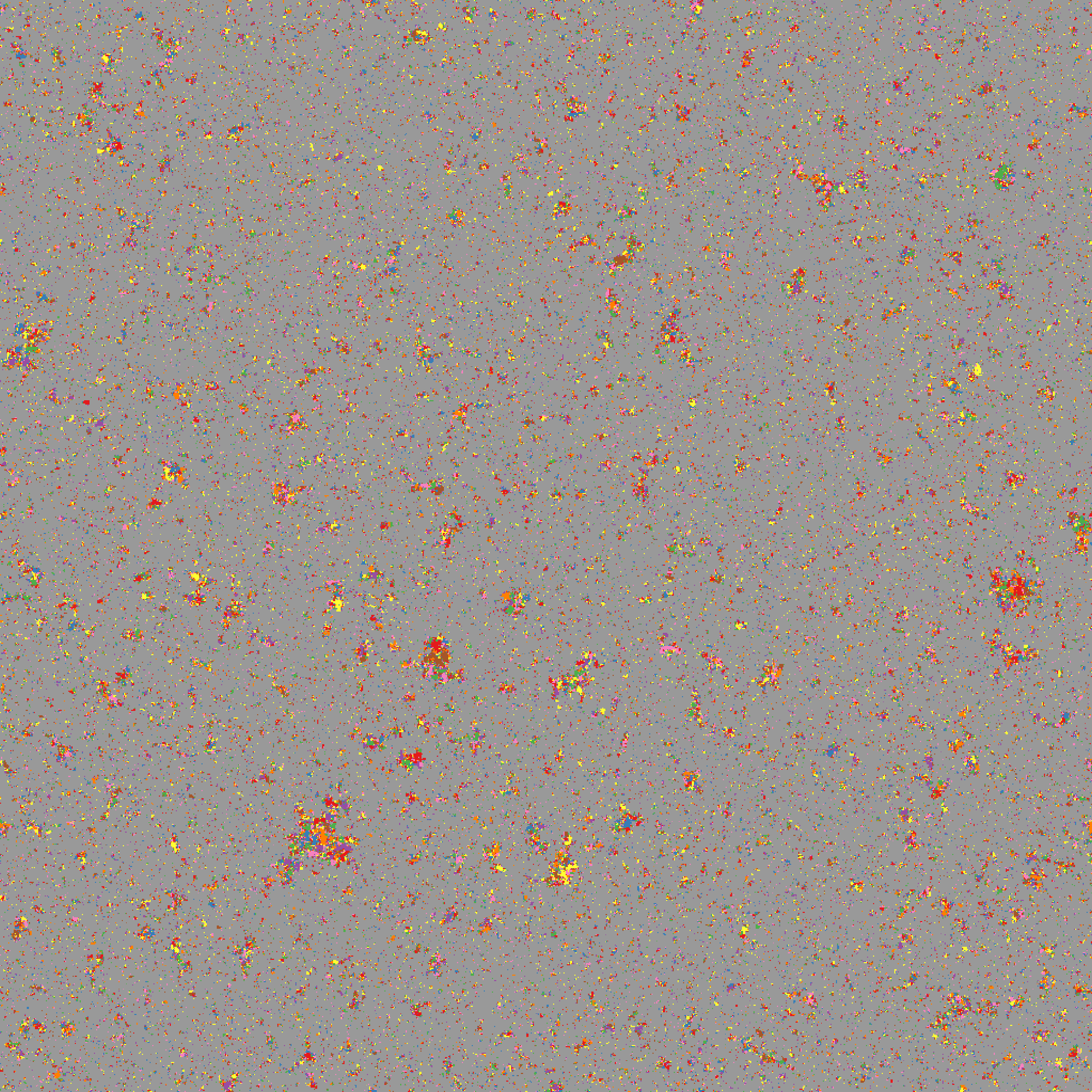}}

	\subcaptionbox{$T=0.724$.}{\includegraphics[scale=0.09]{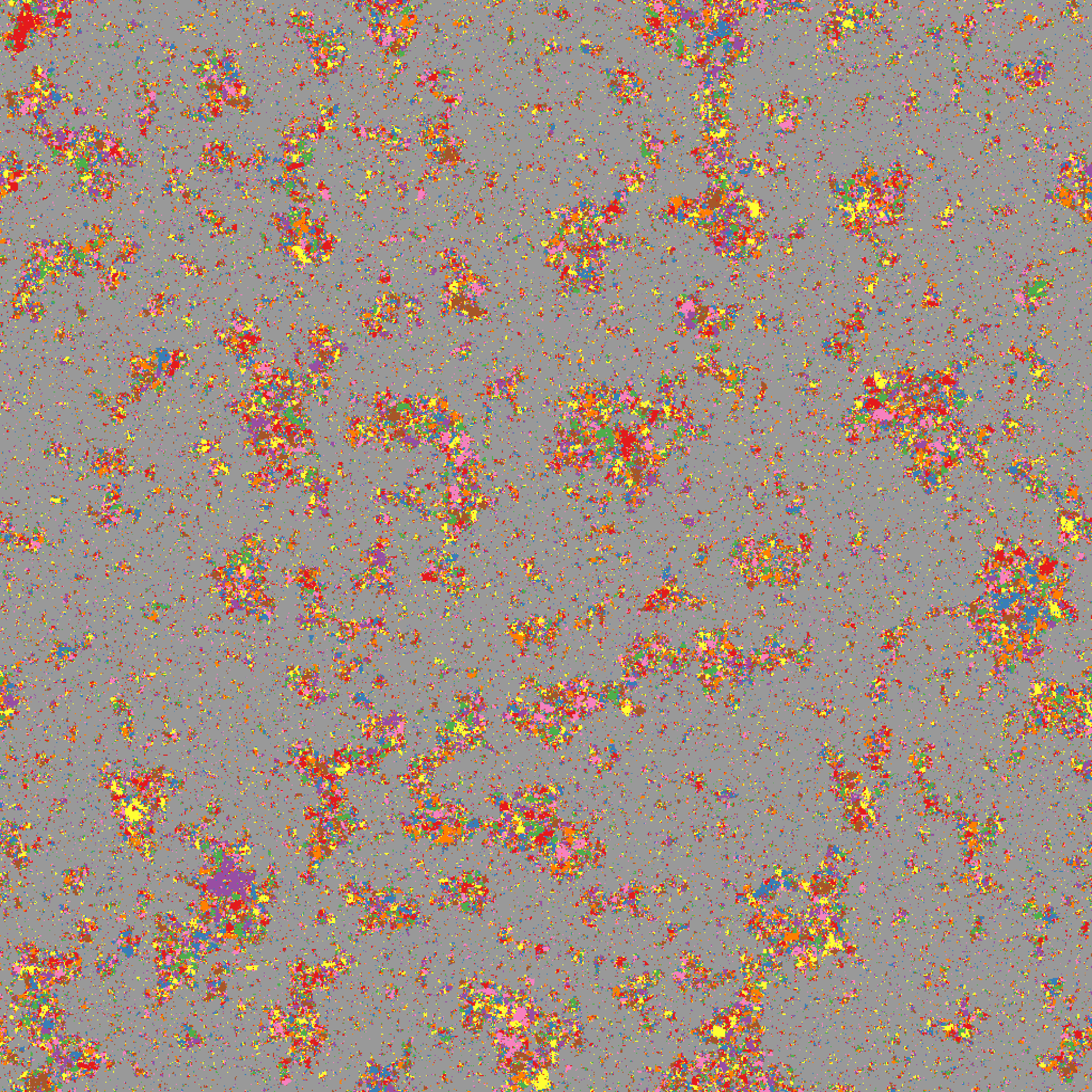}}
	\hfill
	\subcaptionbox{$T=0.73$.}{\includegraphics[scale=0.09]{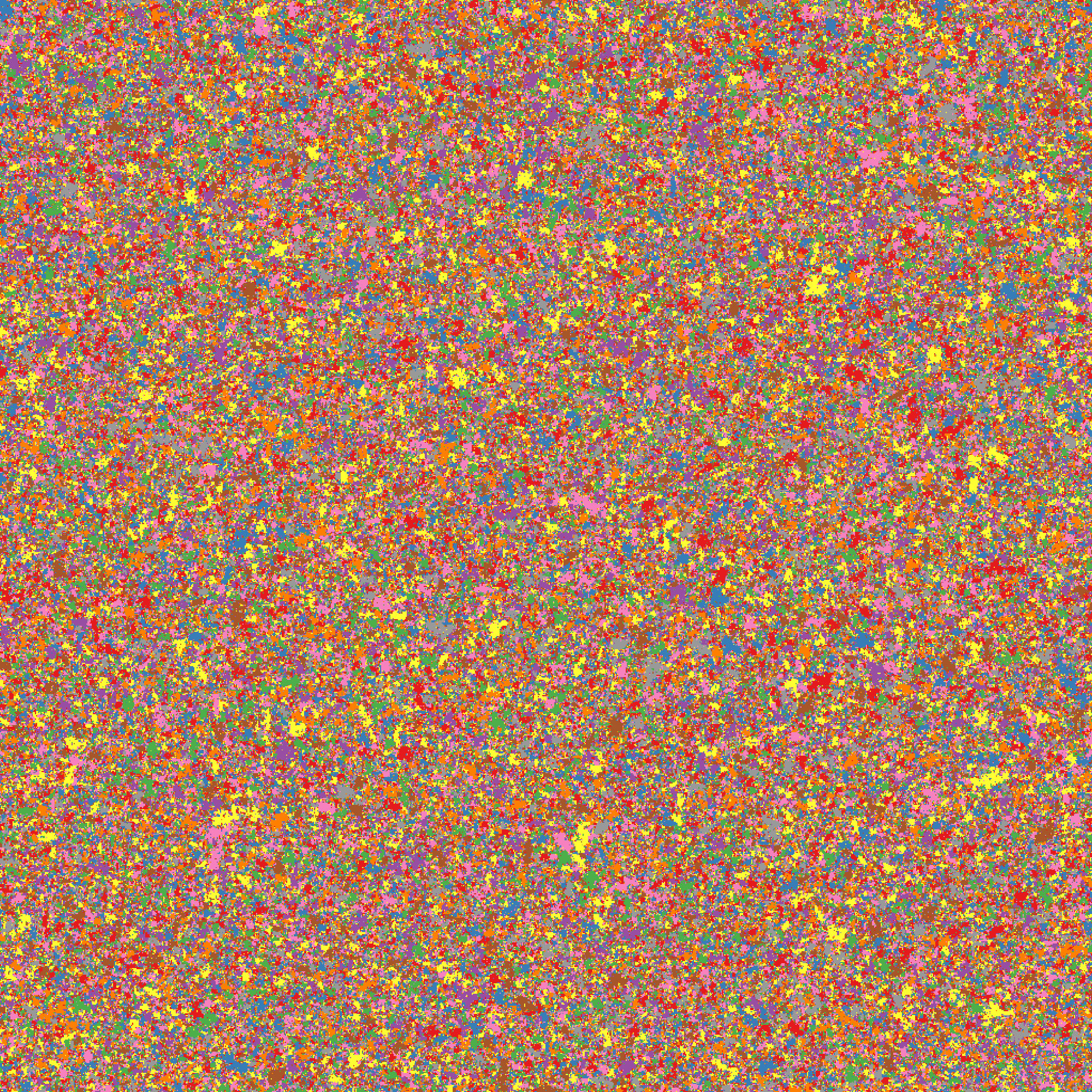}}
	\caption{The $q$-state Potts model rendered with Mìmir at different temperatures ($T$).}
	\label{fig:potts1}
\end{figure}

Listing \ref{code:potts} contains the setup code used to generate the visualizations shown in figure \ref{fig:potts1}. The interop allocations for the color and lattice arrays are added to the code and do not replace previous CUDA allocations. The lattice cells are arranged into an $L \times L$ grid whose positions are generated through the \texttt{makeStructuredGrid} auxiliary method, while the cells themselves hold the spin state indices. The $Q$ state color mapping is achieved by passing the \texttt{colormap} allocation to the color property of the view descriptor and the grid allocation as index, so that each cell value in \texttt{grid} maps to a color stored in \texttt{colormap}.

\begin{code}
\captionof{listing}{Potts model setup code.}
\label{code:potts}
\begin{minted}[linenos,tabsize=2,breaklines]{cpp}
int width = 2000, height = 2000;
createInstance(width, height, &instance);

AllocHandle m1 = nullptr, colormap = nullptr;
allocLinear(instance, (void**)&grid, sizeof(int) * L * L, &m1);

float4 *d_colors = nullptr;
float4 h_colors[Q] = { // Color mapping for the Q states
	{153,153,153,1}, {228,26,28,1}, {55,126,184,1},
	{77,175,74,1}, {152,78,163,1}, {255,127,0,1},
	{255,255,51,1}, {166,86,40,1}, {247,129,191,1},
};
unsigned int num_colors = std::size(h_colors);

auto color_bytes = sizeof(float4) * num_colors;
allocLinear(instance, (void**)&d_colors, color_bytes, &colormap);
CUDA_SAFE_CALL(cudaMemcpy(d_colors, h_colors, color_bytes, cudaMemcpyHostToDevice));

ViewHandle v1 = nullptr;
ViewDescription desc;
desc.element_count = L * L;
desc.extent        = {L, L, 1};
desc.domain_type   = DomainType::Domain2D;
desc.view_type     = ViewType::Voxels;
desc.properties[PropertyType::Position] = makeStructuredGrid(instance, {L,L,1});
desc.properties[PropertyType::Color] = {
	.source  = colormap,
	.size    = num_colors,
	.format  = FormatDescription::make<float4>(),
	.indices = m1,
	.index_size = sizeof(int),
};
createView(instance, &desc, &v1);

displayAsync(instance);
\end{minted}
\end{code}

The visualized interop buffer is mapped to the \texttt{grid} array that stores the full lattice, whose display logic at each timestep is shown at listing \ref{code:potts2}. The \texttt{grid} array is written by \texttt{updateCUDA} from the \texttt{white} and \texttt{black} arrays, updated by the respective \texttt{updateCUDA} kernel calls. The individual packed arrays are not visualized, meaning \texttt{updateCUDA} may run safely outside the interop critical section. 

\begin{code}
\captionof{listing}{Potts model update loop.}
\label{code:potts2}
\begin{minted}[linenos,tabsize=2,breaklines]{cpp}
// white update, read from black
updateCUDA<<<dimGrid, dimBlock>>>(temp, WHITE, white, black);

// black update, read from white
updateCUDA<<<dimGrid, dimBlock>>>(temp, BLACK, black, white);

prepareViews(instance);
dim3 block2(TILE, TILE);
dim3 grid2(L/TILE, (L/2)/TILE);
writeGrid<<<grid2, block2>>>(white, black, grid);
updateViews(instance);
\end{minted}
\end{code}

\subsubsection{Gravitational N-Body}

\begin{figure}[ht!] \centering
	\includegraphics[scale=0.096]{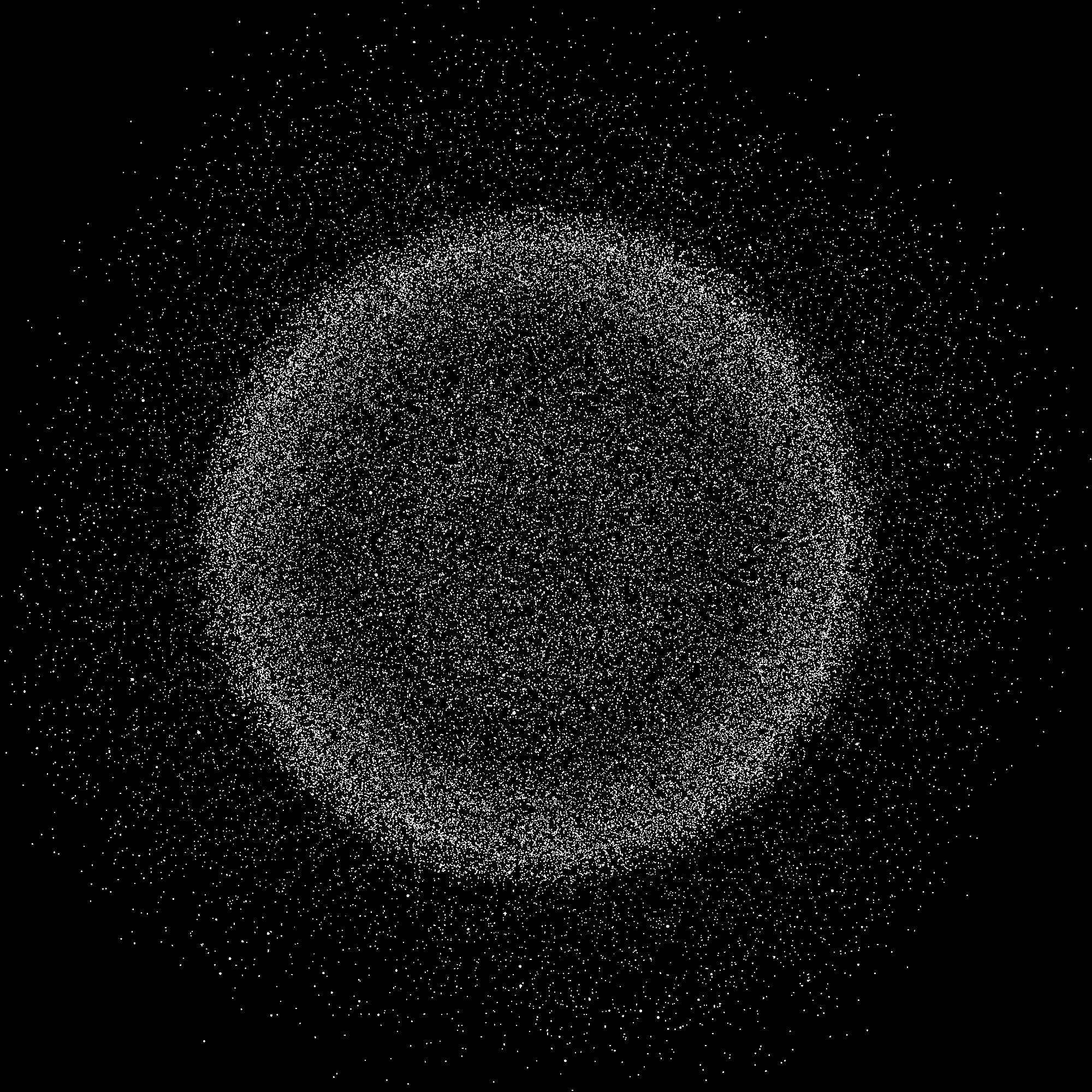}
	\hfill
	\includegraphics[scale=0.096]{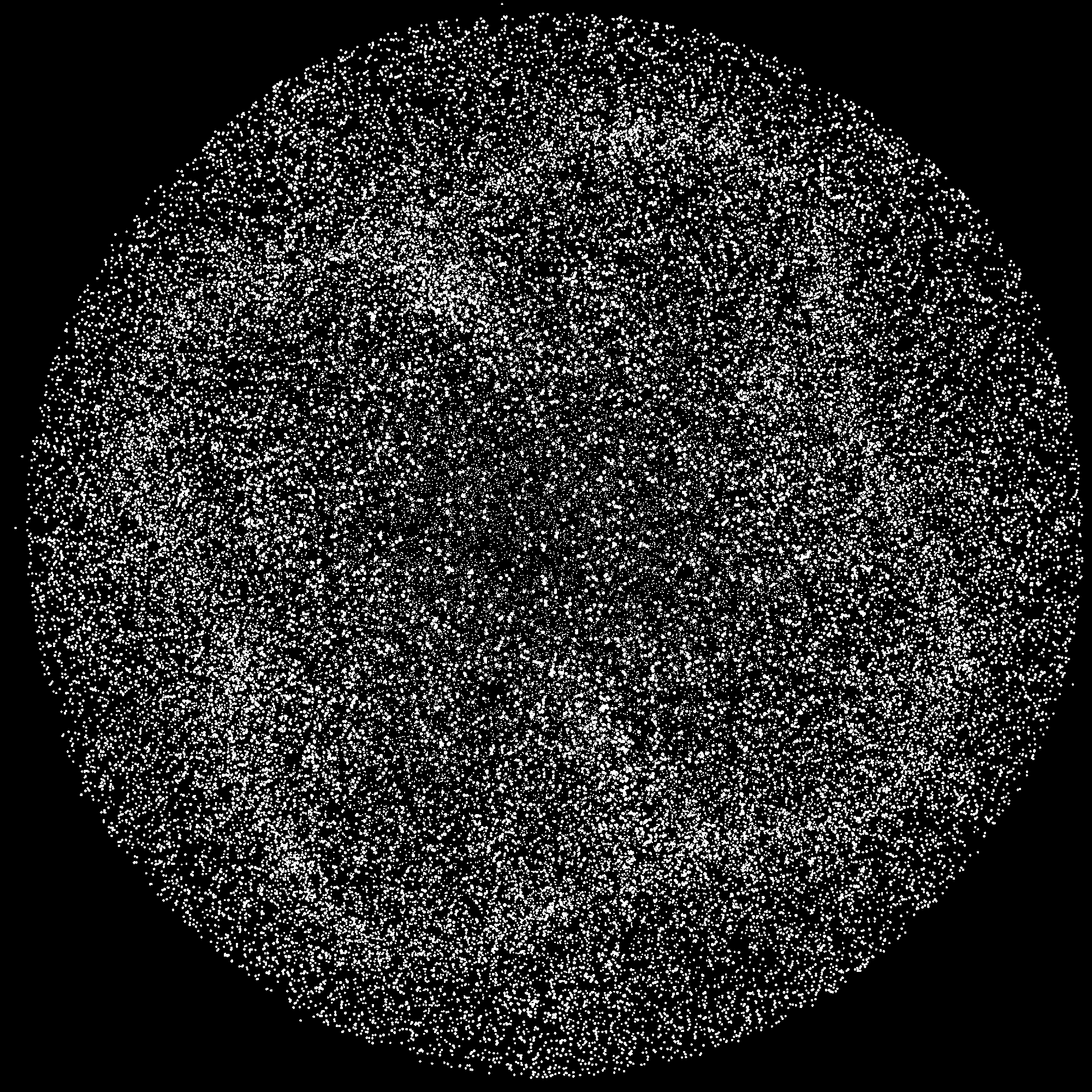}
    
	\includegraphics[scale=0.096]{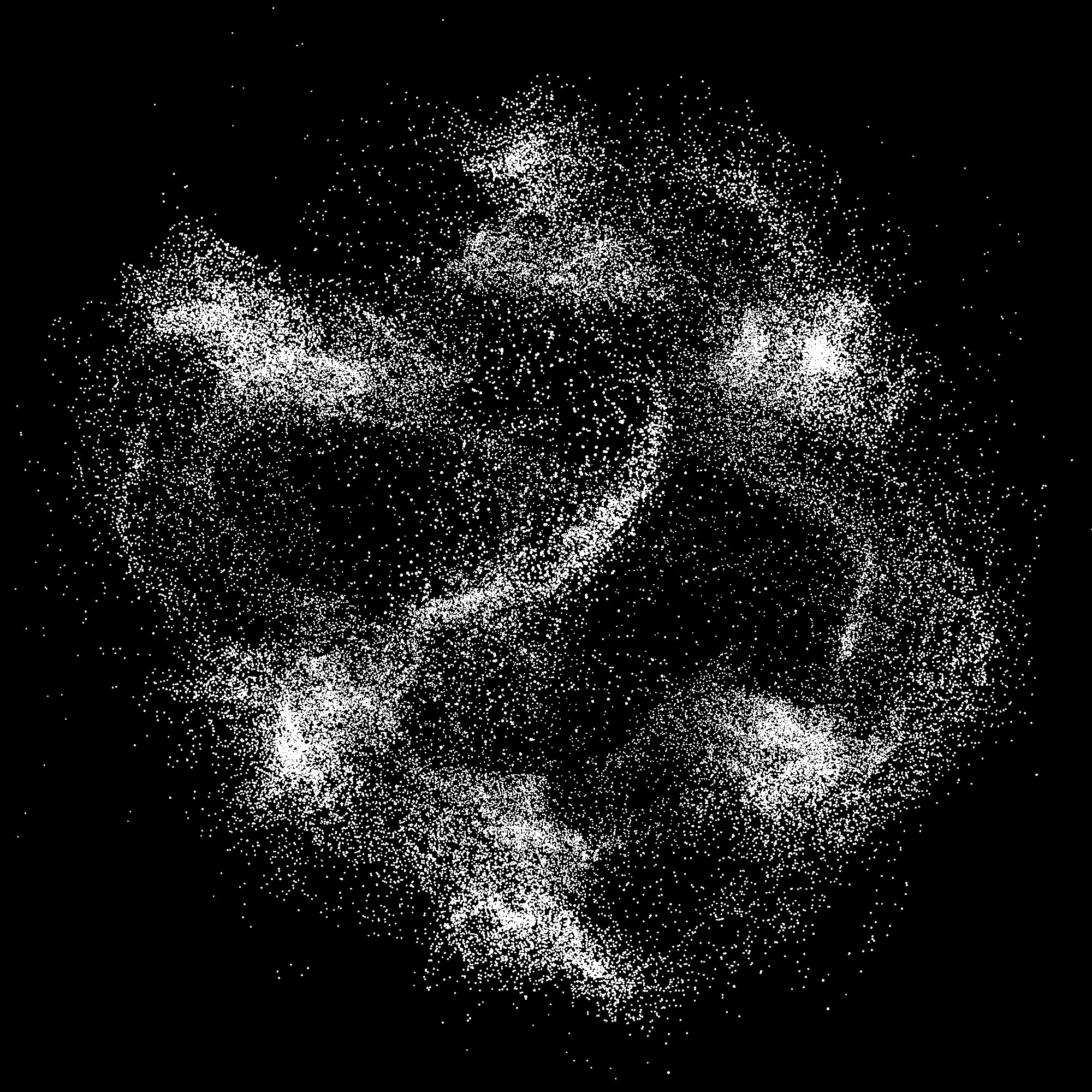}
	\hfill
	\includegraphics[scale=0.096]{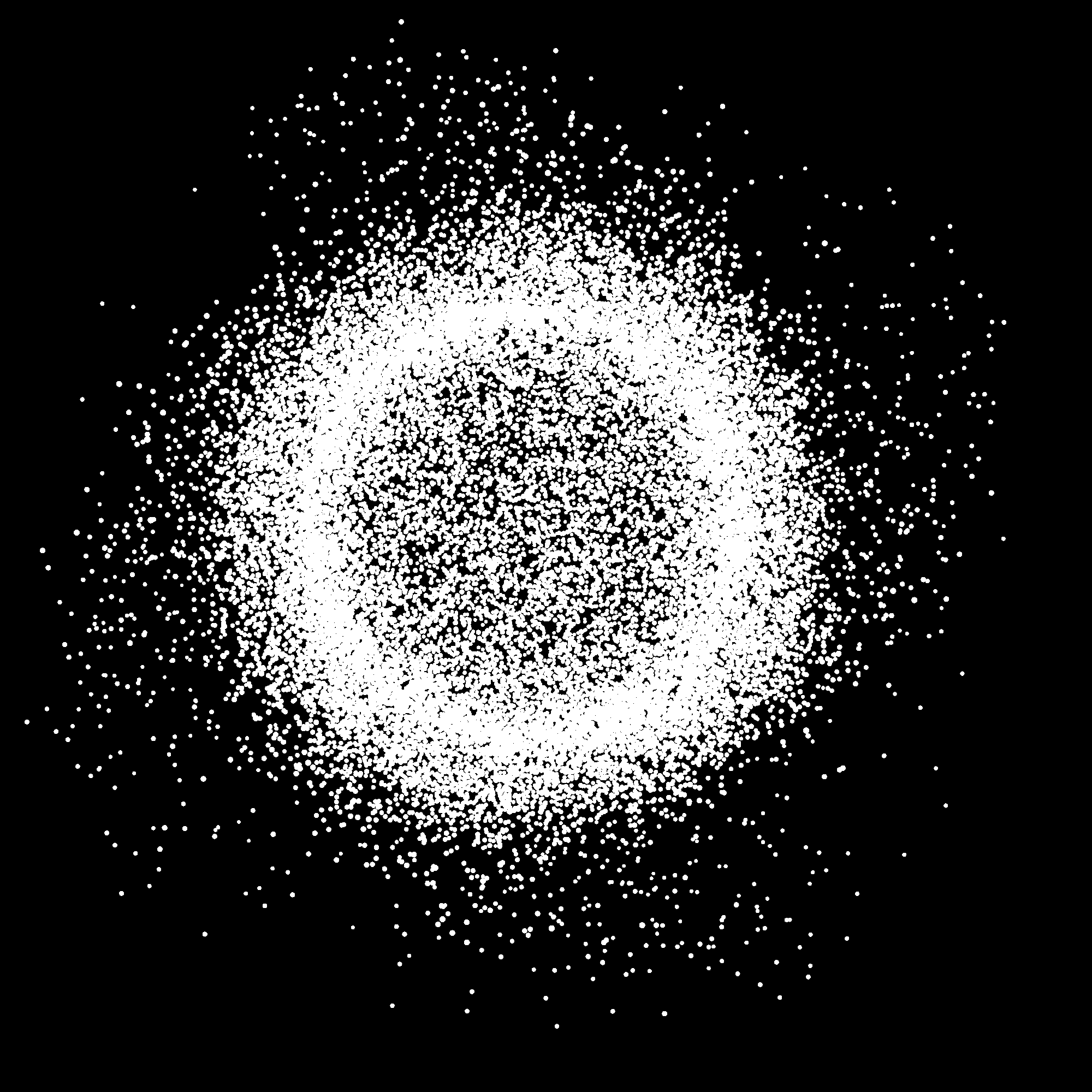}    
	\caption{Gravitational N-Body simulation displayed with Mìmir.}
	\label{fig:grav1}
\end{figure}

This sample shows how an existing visualization code can be ported to use Vulkan as visualization backend through Mìmir, as shown in figure \ref{fig:grav1}. This code is based on the gravitational N-Body program from the Nvidia samples \href{https://github.com/NVIDIA/cuda-samples/tree/master/Samples/5_Domain_Specific/nbody}{repository}. The original code uses CUDA / OpenGL interop to display particles being subject to an all-pairs gravitational force, where the integration step at each iteration is performed by a CUDA kernel call. Encapsulation between experiment and visualization code is improved compared to the original code at the cost of needing to replace CUDA allocation calls with Mìmir linear allocation functions. Listing \ref{code:grav1} shows the setup code for displaying the particle, whose positions use a double-buffered layout (read, write) and a view is created for each. Listing \ref{code:grav2} shows the update step, where the kernel call and read / write buffer swap are from the original code. View visibility is toggled in a likewise fashion to show the updated positions, producing a smooth transition ensured by the critical section between \texttt{prepareViews} and \texttt{updateViews}. The \texttt{isRunning} check is added to the loop condition to exit the simulation when the user closes the display window.

\begin{code}
\captionof{listing}{Gravitational N-body setup code.}
\label{code:grav1}
\begin{minted}[linenos,tabsize=2,breaklines]{cpp}
ViewDescription desc {
    .type   = ViewType::Markers,
    .options = {},
    .domain = DomainType::Domain3D,
    .attributes  = { { AttributeType::Position, {
            .source = allocs[0],
            .size   = input.body_count,
            .format = FormatDescription::make<float4>(),
        }}
    },
    .layout        = Layout::make(input.body_count),
    .visible       = true,
    .default_color = {1.f, 1.f, 1.f, 1.f},
    .default_size  = params.point_size,
    .linewidth     = 0.f,
    .scale         = {1.f, 1.f, 1.f},
};
createView(engine, &desc, &views[0]);

desc.visible = false;
desc.attributes[AttributeType::Position].source = allocs[1];
createView(engine, &desc, &views[1]);
\end{minted}
\end{code}

\begin{code}
\captionof{listing}{Gravitational N-body display code.}
\label{code:grav2}
\begin{minted}[linenos,tabsize=2,breaklines]{cpp}
for (int i = 0; i < input.iter_count && isRunning(engine); ++i) {
    prepareViews(engine);

    integrateNbodySystem(device, current_read, params.time_step,
        params.damping, input.body_count, block_size
    );
    std::swap(current_read, current_write);

    toggleVisibility(views[0]);
    toggleVisibility(views[1]);
    updateViews(engine);
}
\end{minted}
\end{code}

\subsubsection{Triangle meshes}

\begin{figure}[ht!] \centering
	\subcaptionbox{Brain mesh.}{\includegraphics[scale=0.096]{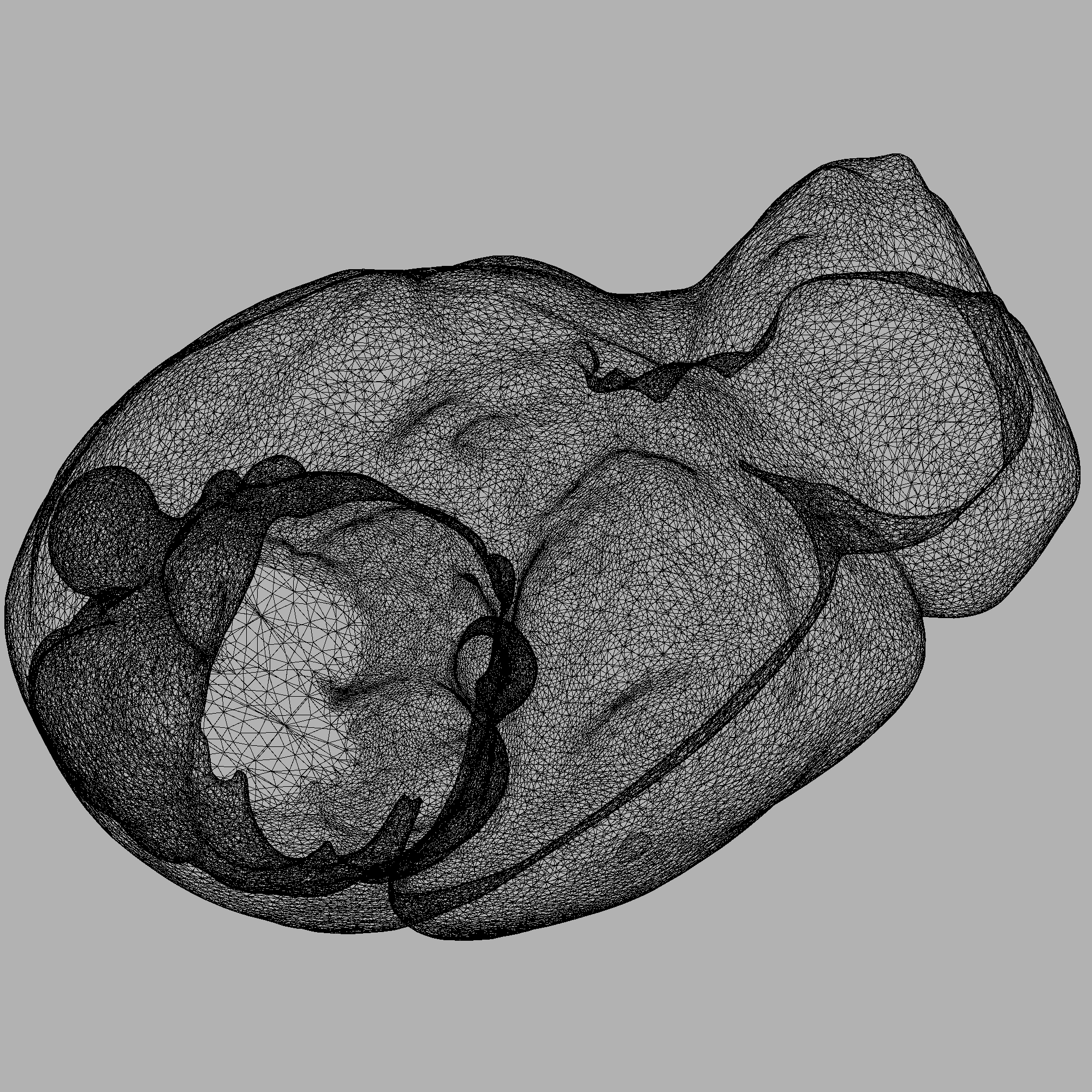}}
	\hfill
	\subcaptionbox{Dragon mesh.}{\includegraphics[scale=0.096]{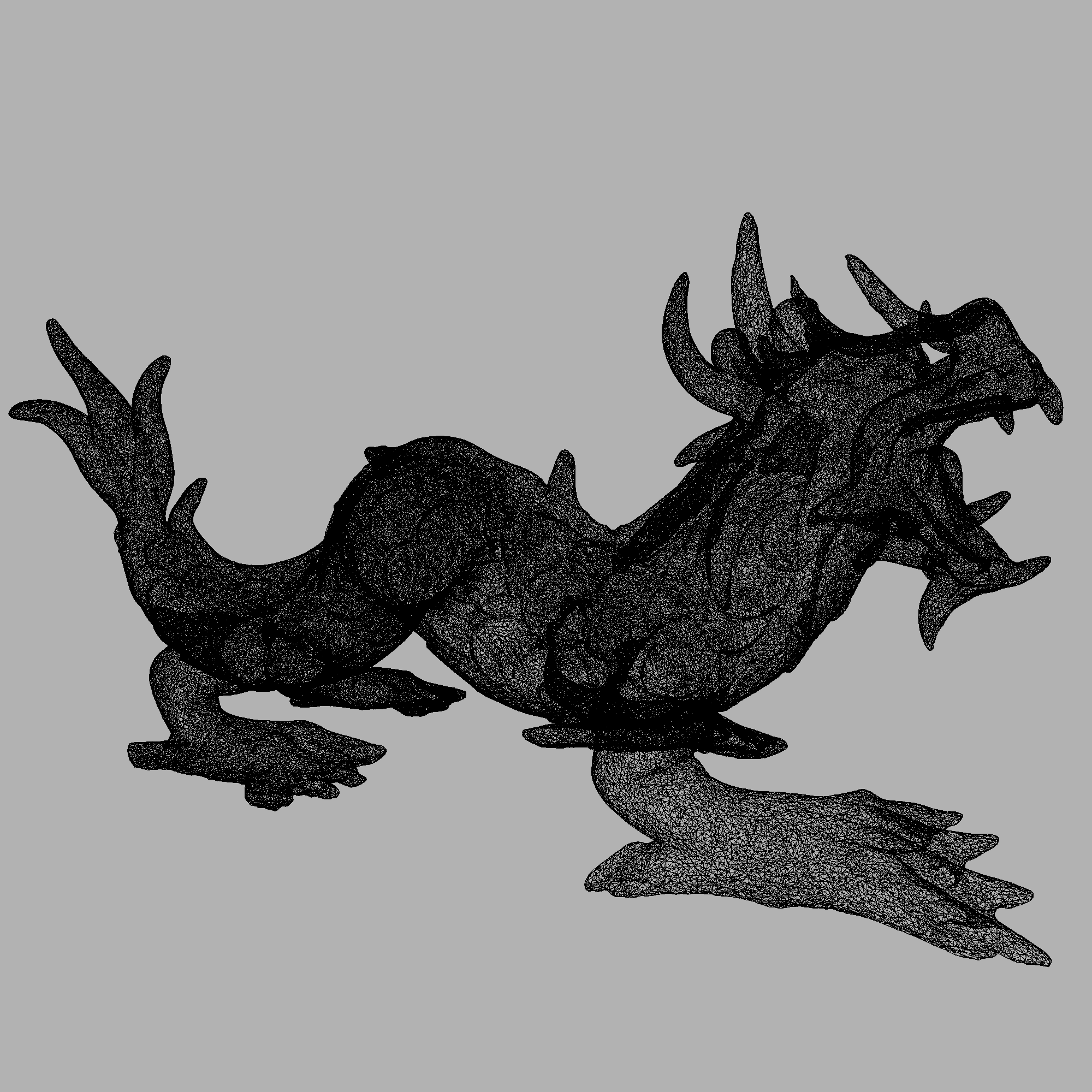}}
	\caption{Mesh data available as input for CUDA kernel execution.}
	\label{fig:mesh}
\end{figure}

Experiments using triangle meshes have a variety of applications in computational physics, such as in Finite Element Analysis (FEA), Computational Fluid Dynamics (CFD), Soft-body dynamics, fluid surfaces, among many others. Unlike previous samples that are based on existing experiments, this shows how to describe the visual attributes of a triangle mesh stored in an interop. The program loads a triangle mesh from an \texttt{.obj} file, computes smooth normals over each of its vertices, and a CUDA kernel deforms the mesh periodically along the normal directions, performing a \say{breathing} motion. Figure \ref{fig:mesh} shows how the mesh is displayed using Mìmir, using views for rendering mesh edges and vertices.

Listings \ref{code:mesh1} and \ref{code:mesh2} shows the setup and display code respectively. Note that the \texttt{vertices} allocation is used in two views simultaneously: in \texttt{v1} as source and in \texttt{v2} as source to be indexed by the content in \texttt{edges}. The loaded vertex coordinates, triangle indices and computed normals are loaded to interop-mapped memory after creating the views, but this can be done as soon as the allocations are complete. The kernel deforming the mesh runs indefinitely until the display window is closed.

\begin{code}
\captionof{listing}{Mesh setup code.}
\label{code:mesh1}
\begin{minted}[linenos,tabsize=2,breaklines]{cpp}
AllocHandle vertices, edges;
auto vert_size = sizeof(float3) * vertex_count;
auto edge_size = sizeof(uint) * triangle_count;
allocLinear(engine, (void**)&d_coords, vert_size, &vertices);
allocLinear(engine, (void**)&d_triangles, edge_size, &edges);

ViewHandle v1 = nullptr, v2 = nullptr;
ViewDescription desc{
    .type   = ViewType::Markers,
    .domain = DomainType::Domain3D,
    .layout = Layout::make(vertex_count),
};
desc.attributes[AttributeType::Position] = {
    .source = vertices,
    .size   = vertex_count,
    .format = FormatDescription::make<float3>(),
};
desc.default_size = 1.f;
desc.linewidth    = 0.f;
createView(engine, &desc, &v1);

// Reuse the above parameters, changing only what is needed
desc.layout = Layout::make(triangle_count);
desc.type = ViewType::Edges;
desc.attributes[AttributeType::Position] = {
    .source   = vertices,
    .size     = vertex_count,
    .format   = FormatDescription::make<float3>(),
    .indexing = {
        .source     = edges,
        .size       = triangle_count,
        .index_size = sizeof(uint32_t),
    }
};
createView(engine, &desc, &v2);

checkCuda(cudaMalloc(&d_normals, vert_size));
checkCuda(cudaMemcpy(d_coords, mesh.vertices.data(), vert_size, cudaMemcpyHostToDevice));
checkCuda(cudaMemcpy(d_triangles, mesh.triangles.data(), edge_size, cudaMemcpyHostToDevice));
checkCuda(cudaMemcpy(d_normals, mesh.normals.data(), vert_size, cudaMemcpyHostToDevice));
\end{minted}
\end{code}

\begin{code}
\captionof{listing}{Mesh setup code.}
\label{code:mesh2}
\begin{minted}[linenos,tabsize=2,breaklines]{cpp}
while (isRunning(engine))
{
    prepareViews(engine);
    breatheKernel<<< grid_size, block_size >>>(d_coords, d_normals, vertex_count, scale);
    scale = varyAngle(degrees);
    updateViews(engine);
}
\end{minted}
\end{code}

\subsection{Discussion}
\label{sec:discussion-conclusions}
The proposed Mìmir library provides the choice to manage compute-graphics synchronization through automatic or manual synchronization, where the latter enables finer control of visualization behavior at the expense of having to declare critical computing sections explicitly in code. It is also possible to disable synchronization altogether, although it should be done with caution making sure that the rendering loop does not take away the GPU performance from the physics simulation, specially for scenes with several geometrical points/primitives to be rendered at each frame. 

The samples shown demonstrate that by adding a few lines of code is is possible to introduce real-time visualizations on CUDA-based simulation codes, never needing to change the structure of the simulation neither having to copy data from device to host memory at each iteration or time step. The only mandatory replacement consists in allocation calls required by the interop workflow, but it is possible to surmount this issue by creating a visualization-only interop array and writing to it when needed. Though this approach requires additional processing, it is a good alternative when the input data is not in a visualization-ready layout, as shown on the Potts model sample.

Benchmark results show the importance of GPU synchronization in interop visualizations beyond preventing visual artifacts, as it fulfills a critical role in balancing compute and graphics workloads when sharing a single device. The performance hit of having interop synchronization is negligible compared to the utility it provides, as disabling may lead to GPU overuse due to competing processes. Due to the importance of synchronization for this kind of experiments, further customization of sync behavior is planned in the future to allow a finer tuning of the compute/graphics relationship, such as selectively enabling/disabling sync sub-steps and synchronizing every $N$ frames.

Choosing Vulkan as a graphics backend for Mïmir instead of OpenGL presents several advantages coming from a more modern API, including a more straightforward interoperability setup that does not have the disadvantages described in section \ref{sec:intro}. OpenGL can be considered a simpler API compared to Vulkan in terms of lines of code, but this simplicity has the drawbacks of having to deal with the OpenGL state machine and more restrictions when dealing with multithread worflows. Vulkan code verbosity allows to keep track of all handles and dependencies involved in visualization, which can be advantageous for complex setups such as interop visualization. The extra programming effort from using Vulkan can pay off when providing support for multiple use cases, such as the presented library. On the contrary, OpenGL may be a better fit for generating a limited set of visualizations from a single code base.

\section{Conclusions}
The proposed library, named Mìmir, shows the feasibility of providing real-time interactive visualizations in GPU computational physics applications for a variety of use cases, while also encapsulating the complexity of a CUDA/Vulkan interoperability environment from experiment code. The library API for interop allocations and views allows to describe relationships between data and properties of a visualization, as was demonstrated by generating visualizations of existing computational physics experiments of different natures and data structures. This scheme allows to communicate directly with experiment code, with the advantage of allowing faster visualization prototyping and real-time visualization, as shown in the sample visualizations.

The presented library implementation targets CUDA and Vulkan, but its design is general enough to be applied to different combinations of compute and graphics APIs, with the consideration that synchronization must be accounted for when establishing interop between them. Using interop between GPU APIs for visualization has the advantage of instant update between a compute process writing to data and a graphics workflow reading it for display, but requires a large amount of code that may exceed experiment size, making code reuse difficult. By streamlining this process in a library, setup and boilerplate code is encapsulated in function calls.

In terms of limitations, the current implementation of Mìmir can suffer degradation in visualization performance if the physical model requires too much computation, not leaving room for the rendering pipeline, or if the number of graphical primitives to render is too large, although this is a normal behavior in graphics. Aspects identified for improvement include frame presentation, rendering (through optimizing shader code), interop synchronization and graphics resource management. Improvement in any of these areas will benefit all current and future use cases, and performance gains can be redirected to support visual improvements when needed. Another limitation is that currently Mìmir can only interoperate with CUDA code, therefore only NVIDIA GPUs can benefit from real-time rendering of the parallel computation. Future work will indeed extend the library to support AMD and Intel GPUs through interoperability with OpenCL code.

Currently Mìmir only supports in-situ rendering, which is a restrictive subset of all available GPU computing platforms. For example, servers lack display capabilities as they use specialized GPUs that do not have display output. Remote rendering support is a planned feature, using the low-level capabilities provided by the CUDA / Vulkan ecosystem to perform efficient frame data transfer. An execution mode of this kind introduces a rendering overhead that will be especially noticeable in iterative simulations. In turn, this raises the need for implementing more computationally economical alternatives, such as coarse-LOD rendering, to minimize latency and improve performance for larger datasets supported by the more powerful server-based GPUs.

\section*{Acknowledgments} \label{credits}
This work was partially funded by ANID  FONDECYT grant 1221357, ANID FONDECYT grant 1241596 and Scholarship Program / MAGISTER NACIONAL / 2022 - 22221291.

\bibliographystyle{elsarticle-num}
\bibliography{references}

\end{document}